\documentclass[twocolumn, showpacs, preprintnumbers,
nofootinbib, aps, prd, superscriptaddress,
10pt, showkeys, notitlepage, longbibliography]{revtex4-1}

\usepackage[dvips]{graphicx}
\usepackage{bm,latexsym,amsmath,amssymb,amsfonts,times}
\usepackage{color}
\usepackage{natbib}
\usepackage{times}
\usepackage{graphicx}
\usepackage{dcolumn}
\usepackage{mathtools}
\newcommand\Tstrut{\rule{0pt}{2.6ex}}
\newcommand\Bstrut{\rule[-0.9ex]{0pt}{0pt}}

\def\be {\begin{equation}}
\def\ee {\end{equation}}
\def\ba {\begin{eqnarray}}
\def\ea {\end{eqnarray}}

\def\bi {\begin{itemize}}
\def\ei {\end{itemize}}

\def\vp {\varphi}
\def\kms{km$^2$ }

\newcommand\beq{\begin{eqnarray}}
\newcommand\eeq{\end{eqnarray}}
\newcommand\tc{{\rm c}}
\newcommand\ts{{\rm s}}
\newcommand\tb{{\rm b}}
\newcommand\tm{{\rm m}}
\newcommand\tcd{{\rm D}}

\usepackage[linktocpage,breaklinks]{hyperref}
\definecolor{venetianred}{rgb}{0.78, 0.03, 0.08}
\definecolor{grey}{rgb}{0.25, 0.25, 0.28}
\definecolor{darkmidnightblue}{rgb}{0.0, 0.2, 0.4}
\definecolor{egyptianblue}{rgb}{0.06, 0.2, 0.65}
\definecolor{darkblue}{rgb}{0.0, 0.0, 0.55}
\hypersetup{colorlinks,breaklinks,
            linkcolor=darkblue,urlcolor=darkblue,
            anchorcolor=darkblue,citecolor=darkblue}

\def\X5sp{{\rm X}_5}
\def\Y3sp{{\rm Y}_3}
\def\Z3sp{{\rm Z}_3}

\begin{document}

\title{Relativistic stars in scalar-tensor theories with disformal coupling}

\author{Masato Minamitsuji}
\email{masato.minamitsuji@ist.utl.pt}
\affiliation{Departamento de F\'isica, CENTRA, Instituto Superior
T\'ecnico, Universidade de Lisboa, Avenida Rovisco Pais 1,
1049 Lisboa, Portugal}

\author{Hector O. Silva}
\email{hokadad@go.olemiss.edu}
\affiliation{Department of Physics and Astronomy, The University of Mississippi, University, Mississippi 38677, USA}

\begin{abstract}
We present a general formulation to analyze the structure of slowly rotating
relativistic stars in a broad class of scalar-tensor theories with
disformal coupling to matter. Our approach includes theories with generalized
kinetic terms, generic scalar field potentials and contains theories with
conformal coupling as particular limits. In order to investigate how the
disformal coupling affects the structure of relativistic stars, we propose a
minimal model of a massless scalar-tensor theory and investigate in detail how
the disformal coupling affects the spontaneous scalarization of slowly rotating
neutron stars.
We show that for negative values of the disformal coupling parameter between
the scalar field and matter, scalarization can be suppressed, while for
large positive values of the disformal coupling parameter stellar models
cannot be obtained. This allows us to put a mild upper bound on this parameter.
We also show that these properties can be qualitatively understood by linearizing
the scalar field equation of motion in the background of a general-relativistic
incompressible star.
To address the intrinsic degeneracy between uncertainties in the equation-of-state
of neutron stars and gravitational theory, we also show the existence
of universal equation-of-state-independent relations between the moment of inertia
and compactness of neutron stars in this theory. We show that
in a certain range of the theory's parameter space the universal relation largely
deviates from that of general relativity, allowing, in principle, to probe
the existence of spontaneous scalarization with future observations.

\end{abstract}

\date{\today}
\maketitle

\section{Introduction}
\label{sec:intro}


Although Einstein's general relativity (GR) has passed all the experimental tests
of gravity in the weak-field/slow-motion regimes with
flying colors~\cite{Will:2014kxa}, it remains fairly unconstrained
in the strong-gravity regime~\cite{Berti:2015itd} and on
the cosmological scales~\cite{Clifton:2011jh}. The recent observation
of gravitational waves generated during the merger of two black holes (BHs) by
the LIGO/Virgo Collaboration, in accordance with general-relativistic
predictions~\cite{Abbott:2016blz,TheLIGOScientific:2016src},
has offered us a first glimpse of gravity in a fully nonlinear and
highly dynamical regime whose theoretical implications are still being
explored~\cite{Yunes:2016jcc}.
Nevertheless, the pressing issues on understanding the nature of
dark matter and dark energy, the inflationary evolution of the early Universe and
the quest for an ultraviolet completion of GR have served as driving forces
in the exploration of modifications to GR~\cite{Clifton:2011jh,Berti:2015itd}.

In general modifications of GR introduce new gravitational degree(s) of
freedom in addition to the metric tensor and can be described by a
scalar-tensor theory of gravity~\cite{Fujii:2003pa}.
On the theoretical side, scalar-tensor theories should not contain
Ostrogradski ghosts~\cite{Woodard:2015zca}, i.e.
the equations of motion should be written in terms of the second-order
differential equations despite the possible
existence of the higher-order derivative interactions at the action level.
On the experimental/observational side, any extension of GR must
pass all the current weak-field tests which GR has successfully passed. Therefore
realistic modifications of gravity should contain a mechanism to suppress
scalar interactions at small scales~\cite{Vainshtein:1972sx,Brax:2004qh} or
(to be interesting) satisfy weak-field tests, but deviate from GR at some
energy scale. Some models satisfying these requirements belongs to the so-called
Horndeski theory~\cite{Horndeski:1974wa,Deffayet:2009mn,Deffayet:2011gz,Kobayashi:2011nu},
the most general scalar-tensor theory with second-order equations of motion.

In scalar-tensor theories, the scalar field may directly couple to matter,
and hence matter does not follow geodesics associated with the metric $g_{\mu\nu}$
but with another ${\tilde g}_{\mu\nu}$. In the simplest case these
two metrics are related as
\begin{eqnarray}
{\tilde g}_{\mu\nu}= A^2(\vp) g_{\mu\nu}\,,
\label{eq:conformal}
\end{eqnarray}
which is known as the conformal coupling \cite{Clifton:2011jh}. The
two frames described by $g_{\mu\nu}$ and ${\tilde g}_{\mu\nu}$
are often referred to as the Einstein and Jordan frames, respectively.

\subsection{Spontaneous scalarization}

For relativistic stars, such as neutron stars (NSs), the conformal
coupling to matter can trigger a tachyonic instability (due to a negative
effective mass) of the scalar field when the star has a
compactness above a certain threshold. This instability
{\it spontaneously scalarizes} the NS, whereupon it harbors a nontrivial
scalar field configuration which smoothly decays outside the star.
In its simplest realization, scalarization occurs
when the conformal factor in Eq. \eqref{eq:conformal} is chosen as
$A({\vp}) = \exp(\beta_1 \vp^2/2)$, where $\beta_1$ is a free parameter of
the theory and $\vp$ is a massless scalar field. This theory passes all
weak-field tests, but the presence of the scalar field can
significantly modify the bulk properties of NSs, such as masses and radii,
in comparison with GR.
This effect was first analyzed for isolated NSs
by Damour and Esposito-Far\`ese~\cite{Damour:1993hw,Damour:1996ke}.
The properties and observational consequences of this phenomenon
were studied in a number of situations, including
stability~\cite{Harada:1997mr,Chiba:1997ms},
asteroseismology~\cite{Sotani:2004rq,Sotani:2005qx,Sotani:2014tua,Silva:2014ora},
slow (and rapidly) rotating NS solutions~\cite{Damour:1996ke,Sotani:2012eb,Doneva:2013qva,Doneva:2014faa,Pani:2014jra},
its influence on geodesic motion of particles
around NSs~\cite{DeDeo:2004kk,Doneva:2014uma},
tidal interactions~\cite{Pani:2014jra} and the
multipolar structure of the spacetime~\cite{Pappas:2014gca,Pappas:2015npa}.
Moreover, the dynamical process of scalarization was studied in
Ref.~\cite{Novak:1998rk} and stellar collapse (including the associated process
of scalar radiation emission) was investigated in
Refs.~\cite{Harada:1996wt,Novak:1997hw,Gerosa:2016fri}.
We refer the reader to Ref.~\cite{Horbatsch:2010hj} for an extensive
literature review.

Additionally, a semiclassical version of this effect~\cite{Lima:2010na}
(cf. also~\cite{Landulfo:2012nz,Mendes:2013ija,Landulfo:2014wra,Mendes:2014vna}
and~\cite{Pani:2010vc} for a connection with the Damour-Esposito-Far\`ese
model~\cite{Damour:1993hw,Damour:1996ke})
has been shown to {\it awaken the vacuum} state of a quantum field
leading to an exponential growth of its vacuum energy density in the background
of a relativistic star.

These nontrivial excitations of scalar fields induced by relativistic stars are a
consequence of the generic absence of a ``no-hair theorem'' for these objects
(see Refs.~\cite{Yagi:2011xp,Yagi:2015oca,Barausse:2015wia} for counterexamples),
in contrast to the case of BHs,
and can potentially be an important source for signatures of the presence of
fundamental gravitational scalar degrees of freedom
through astronomical observations~\cite{Psaltis:2008bb,Berti:2015itd},
including the measurements of gravitational and
scalar radiation signals~\cite{Yunes:2013dva}.

The phenomenological implications of spontaneous scalarization have also
been explored in binary NS
mergers~\cite{Barausse:2012da,Palenzuela:2013hsa,Taniguchi:2014fqa,Sennett:2016rwa}
and in BHs surrounded by matter~\cite{Cardoso:2013fwa,Cardoso:2013opa}. In
the former situation, a {\it dynamical} scalarization allows binary members to
scalarize under conditions where this would not happen if they were isolated. This
effect can dramatically change the dynamics of the system in the final cycles
before the merger with potentially observable consequences.
In the latter case, the presence of matter can cause the
appearance of a nontrivial scalar field configuration, growing ``hair''
on the BH.

On the experimental side, binary-pulsar observations~\cite{Freire:2012mg}
have set stringent bounds on $\beta_1$, whose value is presently constrained to
be $\beta_1 \gtrsim -4.5$. This tightly constrains the effects of spontaneous
scalarization in isolated NSs, for it has been shown that independently of the
choice of the equation of state (EOS) scalarization can occur only if
$\beta_1 \lesssim -4.35$ for NSs modeled by a perfect
fluid~\cite{Harada:1998ge,Novak:1998rk,Silva:2014fca}. These two results
confine $\beta_1$ to a very limited range, in which, even if it exists in
nature, the effects of scalarization on isolated NSs are bound to be small;
see Refs.~\cite{Silva:2014fca,Doneva:2013qva} for examples where the
threshold value of $\beta_1$ can be increased and
Refs.~\cite{Mendes:2014ufa,Palenzuela:2015ima,Mendes:2016fby}
for recent work exploring the large positive $\beta_1$ region of the theory.

\subsection{Disformal coupling}
\label{sec:disformalcoupling}

It was recently understood that modern scalar-tensor theories of gravity,
under the umbrella of Horndeski gravity~\cite{Horndeski:1974wa,Deffayet:2009wt},
offer a more general class of coupling~\cite{Bettoni:2013diz,Zumalacarregui:2013pma}
between the scalar field and matter through the so-called
{\it disformal coupling}~\cite{Bekenstein:1992pj}
\begin{eqnarray}
\label{eq:disformal}
{\tilde g}_{\mu\nu}=A^2 (\vp)
\left[g_{\mu\nu}+ \Lambda B^2(\vp)\vp_{\mu}\vp_{\nu}\right],
\end{eqnarray}
where $\vp_\mu = \nabla_\mu\varphi$ is the covariant derivative of the scalar
field associated with the gravity frame metric $g_{\mu\nu}$,
and $\Lambda$ is a constant with dimensions of $({\rm length})^2$.
For $\Lambda = 0$ we recover the purely conformal case of Eq.~\eqref{eq:conformal}.
Disformal transformations were originally introduced by Bekenstein and consist
of the most general coupling constructed from the metric $g_{\mu\nu}$ and the
scalar field $\varphi$ that respects causality and the weak equivalence
principle~\cite{Bekenstein:1992pj}.
Disformal couplings have been investigated so far mainly in the context of
cosmology~\cite{Koivisto:2008ak,Sakstein:2014aca,Sakstein:2015jca}.
They also arise in higher-dimensional gravitational theories
with moving branes \cite{Zumalacarregui:2012us,Koivisto:2013fta}
in relativistic extensions of modified Newtonian theories,
the tensor-vector-scalar theories~\cite{Bekenstein:2004ne,Bekenstein:1993fs},
and
in the decoupling limit of the nonlinear massive gravity \cite{deRham:2010kj,deRham:2010ik,Berezhiani:2013dw,Brito:2014ifa}.
Moreover, in Ref.~\cite{Bettoni:2013diz} it was shown that
the mathematical structure of Horndeski theory is
preserved under the transformation \eqref{eq:disformal},
namely if the scalar-tensor theory written in terms of $g_{\mu\nu}$
belongs to a class of the Horndeski theory
the same theory rewritten in terms of ${\tilde g}_{\mu\nu}$
belongs to another class of the Horndeski theory.
Thus disformal transformations provide a natural generalization of
conformal transformations.

Disformal coupling was also considered in models of a varying speed of
light~\cite{Magueijo:2003gj} and inflation~\cite{Kaloper:2003yf,vandeBruck:2015tna}.
The invariance of cosmological observables in the frames related by the
disformal relation~\eqref{eq:disformal} was verified
in Refs.~\cite{Creminelli:2014wna,Minamitsuji:2014waa,Tsujikawa:2014uza,Watanabe:2015uqa,Motohashi:2015pra,Domenech:2015hka}.
Although applications to early Universe models are still limited, disformal
couplings have been extensively applied to late-time cosmology~\cite{Sakstein:2014isa,vandeBruck:2015ida,Sakstein:2015jca,Koivisto:2008ak,Zumalacarregui:2012us,Zumalacarregui:2010wj,Koivisto:2012za,DeFelice:2011bh,Bettoni:2012xv,Hagala:2015paa}.
A new screening mechanism of the scalar force in the high-density
region was proposed in Ref.~\cite{Koivisto:2012za}, where in the presence
of disformal coupling the nonrelativistic limit of the scalar field equation
seemed to be independent of the local energy density. However, a reanalysis
suggested that no new screening mechanism from disformal coupling
could work \cite{Sakstein:2014isa,Ip:2015qsa}.
It was also argued that disformal coupling could not contribute to a chameleon screening mechanism around a nonrelativistic source \cite{Noller:2012sv}.
Experimental and observational constraints on disformal coupling to particular
matter sectors have also been investigated. Disformal couplings to baryons
and photons have been severely constrained in terms of
the nondetection of new physics in collider experiments
~\cite{Kaloper:2003yf,Brax:2014vva,Brax:2014zba,Brax:2014vla,Lamm:2015gka,Brax:2015hma},
the absence of spectral distortion of the cosmic microwave background
and the violation of distance reciprocal relations
~\cite{vandeBruck:2013yxa,Brax:2013nsa,Brax:2014vla,vandeBruck:2015rma},
respectively. On the other hand, disformal coupling to the dark sector has
been proposed in~\cite{Neveu:2014vua,vandeBruck:2015ida}
and is presently less constrained in comparison with coupling to
visible matter sectors.

When conformal and disformal couplings are {\it universal}
to all the matter species, they can only be constrained
through experimental tests of gravity.
A detailed study of scalar-tensor theory with the pure disformal coupling
$A(\varphi)=1$ and $B(\varphi)=1$ in the weak-field limit was
presented in~\cite{Sakstein:2014isa} and the post-Newtonian (PN) corrections due to
the presence of pure disformal coupling were computed~\cite{Ip:2015qsa}.
In these papers~\cite{Sakstein:2014isa,Ip:2015qsa},
in contrast to the claim of Refs.~\cite{Zumalacarregui:2012us,Koivisto:2012za},
it was shown that no screening mechanism which could suppress the scalar force
in the vicinity of the source exists and the difference of the parametrized
post-Newtonian (PPN) parameters from GR are of order $|\Lambda|H_0^2$,
where $H_0 (\sim 10^{-28}\, {\rm cm}^{-1})$ is the present-day Hubble scale.
The strongest bound on $|\Lambda|$ comes from the constraints on the PPN preferred
frame parameter $\alpha_2$. The near perfect alignment between the Sun's
spin axis and the orbital angular momenta of the planets provides the
constraint $\alpha_2<4 \times 10^{-5}$ (see Ref.~\cite{Iorio:2012pe} for
a discussion), which implies that
$|\Lambda|\lesssim 10^{-6} H_0^{-2}(\sim 10^{40}\, {\rm km}^2)$.
With the inclusion of the conformal factor, i.e. $A(\varphi)\neq 1$,
the authors of Ref.~\cite{Ip:2015qsa} argued that the Cassini bound
$|\gamma-1|<2.1\times 10^{-5}$ \cite{Bertotti:2003rm}
imposes a constraint on $\alpha(\varphi_0)$, where $\vp_0$ is the
cosmological background value of the scalar field and
\begin{equation}
\label{eq:alphabeta}
\alpha(\varphi) \coloneqq \frac{d\log A(\varphi)}{d\varphi},
\quad
\beta(\varphi) \coloneqq \frac{d\log B(\varphi)}{d\varphi}.
\end{equation}
On the other hand the disformal part of the coupling $\beta(\varphi_0)$
remains unconstrained, because corrections to the PPN parameters which include
$\beta(\vp)$ are subdominant compared to the conformal part.
These weaker constraints on the disformal coupling
parameters are due to the fact that in the nonrelativistic regime
with negligible pressure and a slowly evolving scalar field the disformal
coupling becomes negligible. We also point out that in the weak-field
regime such as in the Solar System, typical densities are small therefore
preventing the appearance of ghosts in the theory for negative values of
$\Lambda$.

In the strong-gravity regime such as
that found in the interior of NSs, the pressure cannot be neglected
and the disformal coupling is expected to be as important as the conformal one.
This would affect the spontaneous scalarization mechanism and
consequently influence the structure (and stability) of relativistic stars,
or have significant impact on gravitational-wave
astronomy~\cite{Berti:2015itd}.
The influence of disformal coupling on the stability of matter configurations
around BHs was analyzed in Ref.~\cite{Koivisto:2015mwa}.
The authors of Ref.~\cite{Koivisto:2015mwa} derived
the stability conditions of the system by generalizing the case of
pure conformal coupling~\cite{Cardoso:2013fwa,Cardoso:2013opa}.
They also generalized these works to scalar-tensor theories with noncanonical
kinetic terms and disformal coupling, finding that the
disformal coupling could make matter configurations more unstable,
triggering spontaneous scalarization. In the present work
within the same class of scalar-tensor theory considered in
Ref.~\cite{Koivisto:2015mwa}, we will study relativistic stars and
investigate the influence of disformal coupling on the scalarization of NSs.

\subsection{Organization of this work}
\label{sec:organization}
This paper is organized as follows.
In Sec.~\ref{sec:theory} we review the fundamentals of
scalar-tensor theories with generalized kinetic term and disformal coupling.
In Sec.~\ref{sec:structure} we present a general formulation to analyze the
structure of slowly rotating stars in theories with disformal coupling.
In Sec.~\ref{sec:case}, as a case study, we consider a canonical scalar field
with a generic scalar field potential. We particularize the stellar
structure equations to this model and discuss how to solve them numerically.
In Sec.~\ref{sec:incompress} we explore the consequences of the disformal
coupling by studying small scalar perturbations to an incompressible relativistic
star in GR. In particular we investigate the conditions for which spontaneous
scalarization happens.
In Sec.~\ref{sec:results} we present our numerical studies about the influence of
disformal coupling on the spontaneous scalarization by solving the full
stellar structure equations.
In Sec.~\ref{sec:icrelations} as an application of our numerical integrations,
we examine the EOS independence between the moment of inertia and compactness
of NSs in scalar-tensor theory comparing it against the results obtained in GR.
Finally, in Sec.~\ref{sec:conclusions} we summarize our main findings and
point out possible future avenues of research.

\section{Scalar-tensor theory with the disformal coupling}
\label{sec:theory}

We consider scalar-tensor theories in which matter is disformally coupled to
the scalar field. The action in the Einstein frame reads
\begin{align}
S &= \frac{1}{2\kappa} \int d^4x \sqrt{-g}
\left[
R +2 P(X,\varphi)
\right] \nonumber \\
&+\int d^4x\sqrt{-\tilde g\left(\varphi,\varphi_\mu\right)}\,
{\cal L}_\tm\left[{\tilde g}_{\mu\nu}\left(\varphi,\varphi_\mu\right),\Psi\right],
\label{eq:e_ac}
\end{align}
where
$x^{\mu}$ ($\mu=0,1,2,3$)
 represents the coordinate system of the spacetime,
$g_{\mu\nu}$ and  ${\tilde g}_{\mu\nu}$ are respectively the Einstein and Jordan frame metrics
disformally related by \eqref{eq:disformal},
$g:={\rm det} (g_{\mu\nu})$ and ${\tilde g}:={\rm det} ({\tilde g}_{\mu\nu})$,
$R$ is the Ricci scalar curvature associated with $g_{\mu\nu}$,
$\kappa \coloneqq (8\pi G)/c^4$, where $G$ is the gravitational constant defined in
the Einstein frame and $c$ is the speed of light in vacuum.
$P(X,\varphi)$ is an arbitrary function of
the scalar field $\varphi$
and $X\coloneqq-\frac{1}{2}g^{\mu\nu}\varphi_\mu\varphi_\nu$,
and
${\cal L}_\tm$ represents the Lagrangian density of matter fields $\Psi$.
We note that the canonical scalar field corresponds to the
case of $P(X,\varphi)=2X -V(\varphi)$, but we will not restrict the form of
$P(X,\varphi)$ at this stage.
In this paper we will not omit $G$ and $c$.

Varying the action \eqref{eq:e_ac} with respect to the Einstein frame metric
$g_{\mu\nu}$, we obtain the Einstein field equations
\begin{equation}
G^{\mu\nu}=\kappa \left(T_{(\tm)}^{\mu\nu}+ T_{(\varphi)}^{\mu\nu}\right),
\label{eq:eins}
\end{equation}
where the energy-momentum tensors of the matter fields $\Psi$ and scalar field
$\varphi$ are given by
\begin{align}
T_{(\tm)}^{\mu\nu}
=\frac{2}{\sqrt{-g}}
\frac{\delta \left(\sqrt{-\tilde g} {\cal L}_\tm\left[{\tilde g}(\varphi),\Psi\right]\right)}
{\delta g_{\mu\nu}},
\end{align}
and
\begin{align}
T_{(\varphi)}^{\mu\nu} &\coloneqq
\frac{1}{\kappa}
\frac{2}{\sqrt{-g}}
\frac{\delta \left(\sqrt{-g} P(X,\varphi)\right)}
{\delta g_{\mu\nu}}
\nonumber \\
&=\frac{1}{\kappa}
\left(
P_X\varphi^\mu\varphi^\nu
+
P g^{\mu\nu}
\right),
\end{align}
respectively, where $P_X\coloneqq{\partial_{X} P}$
and $\varphi^\mu:=g^{\mu\nu}\varphi_\nu$.
From Eq.~\eqref{eq:disformal}, the inverse Jordan frame metric ${\tilde g}^{\mu\nu}$ is
related to the inverse Einstein frame metric $g^{\mu\nu}$ by
\begin{equation}
\label{eq:inv_disformal}
{\tilde g}^{\mu\nu}
=A^{-2}(\varphi)
\left[g^{\mu\nu}
-\frac{\Lambda B^{2}(\varphi)}{\chi (X,\varphi)}
\varphi^\mu
\varphi^\nu
\right],
\end{equation}
where we have defined
\begin{equation}
\label{eq:chi}
\chi (X,\varphi)\coloneqq 1-2\Lambda B^2(\varphi) X.
\end{equation}
The volume element in the Jordan frame $\sqrt{-\tilde g}$ is given by
$\sqrt{-\tilde g}=A^4(\varphi)\sqrt{-g}\sqrt{\chi (X,\varphi)}$.
In order to keep the Lorentzian signature of the Jordan frame metric
${\tilde g}_{\mu\nu}$, $\chi$ must be non-negative.
We note that in the purely conformal coupling limit
$\Lambda=0$ and $\chi = 1$.

The contravariant energy-momentum tensor in the Jordan frame
${\tilde T}^{\mu\nu}_{(\tm)}$ is related to that in the Einstein frame by
\begin{align}
{\tilde T}_{(\tm)}^{\mu\nu}
&\coloneqq
\frac{2}{\sqrt{-\tilde g}}
\frac{\delta \left(\sqrt{-\tilde g} {\cal L}_\tm\left[{\tilde g},\Psi\right]\right)}
{\delta \tilde g_{\mu\nu}},
\nonumber \\
&=
\sqrt{\frac{g}{\tilde g}}
\frac{\delta g_{\alpha\beta}}{\delta {\tilde g}_{\mu\nu}}
T_{(\tm)}^{\alpha\beta}
=\frac{A^{-6}(\varphi)}
        {\sqrt{\chi(X,\varphi)}}
T_{(\tm)}^{\mu\nu}.
\end{align}
The mixed and covariant energy-momentum tensors in the Jordan frame are
respectively given by
\begin{subequations}
\begin{align}
\label{mixed1}
{\tilde T}_{(\tm)\mu}{}^{\nu}
&=
\frac{A^{-4}(\varphi)}
        {\sqrt{\chi (X, \varphi)}}
\left(
  \delta_\mu^\alpha
+\Lambda B^{2}(\vp)
 \varphi_\mu\varphi^\alpha
\right)
T_{(m)\alpha}{}^{\nu},
\\
{\tilde T}_{(\tm)\mu\nu}
&=
\frac{A^{-2}(\vp)}
        {\sqrt{\chi (X,\varphi)}}
\left(
  \delta_\mu^\alpha
+\Lambda B^{2}(\varphi)
 \varphi_\mu\varphi^\alpha
\right) \nonumber \\
&\times \left(
  \delta_\nu^\beta
+\Lambda B^{2}(\varphi)
 \varphi_\nu\varphi^\beta
\right)
T_{(\tm)\alpha\beta},
\end{align}
\end{subequations}
and
\begin{subequations}
\begin{align}
T_{(\tm)}^{\mu\nu}
&=
A^{6}(\varphi) \sqrt{\chi (X,\varphi)}
{\tilde T}_{(\tm)}^{\mu\nu},
\\
\label{stu}
T_{(\tm)\nu}{}^{\mu}
&=
A^{4}(\varphi) \sqrt{\chi (X,\varphi)}
\left(
\delta_\nu^\rho
-\frac{\Lambda B^{2}(\varphi)\varphi^\rho\varphi_\nu}{\chi (X,\varphi)}
\right)
{\tilde T}_{(\tm)\rho}{}^{\mu},\\ \nonumber \\
T_{(\tm)\mu\nu}
&=
A^{2}(\varphi)
 \sqrt{\chi  (X,\varphi)}
\left(
\delta_\mu^\rho
-\frac{\Lambda B^{2}(\varphi)\varphi^\rho\varphi_\mu}{\chi (X,\varphi)}
\right)\nonumber \\
&\times\left(
\delta_\nu^\sigma
-\frac{\Lambda B^{2}(\varphi)\varphi^\sigma\varphi_\nu}{\chi (X,\varphi)}
\right)
{\tilde T}_{(\tm)\rho\sigma}.
\end{align}
\end{subequations}

In terms of the covariant tensors, the Einstein equations in the Einstein
frame~\eqref{eq:eins} can be recast as
\begin{align}
\label{eq:einstein_cov}
G_{\mu\nu} &=
\kappa A^{2}(\varphi)
 \sqrt{\chi (X,\varphi) }
\left(
\delta_\mu^\rho
-\frac{\Lambda B^{2}(\varphi)\varphi^\rho\varphi_\mu}{\chi (X,\varphi)}
\right) \nonumber \\
&\times
\left(
\delta_\nu^\sigma
-\frac{\Lambda B^{2}(\varphi)\varphi^\sigma\varphi_\nu}{\chi  (X,\varphi)}
\right)
{\tilde T}_{(\tm)\rho\sigma}
+ P_X\varphi_\mu\varphi_\nu \nonumber \\
&+g_{\mu\nu}P.
\end{align}
Varying the action \eqref{eq:e_ac} with respect to the scalar field $\varphi$,
we obtain the scalar field equation of motion
\begin{equation}
P_X\Box\varphi+ P_\varphi
-P_{XX}\varphi^{\rho}\varphi^\sigma\varphi_{\rho\sigma}-2XP_{X\varphi}
=\kappa {\cal Q},
\label{eq:scalar}
\end{equation}
where the function ${\cal Q}$ characterizes the strength of the coupling of matter
to the scalar field
\begin{align}
\label{eq:coup}
{\cal Q} &\coloneqq
\Lambda
\nabla_\rho
\left(
B^2(\varphi)
T_{(\tm)}^{\rho\sigma}
\varphi_\sigma
\right)
-\alpha (\varphi)
 T_{(\tm)} \nonumber \\
&-\Lambda B^2(\varphi)
\left[
\alpha(\varphi)
+\beta(\varphi)
\right]
 T_{(\tm)}^{\rho\sigma}
 \varphi_\rho
 \varphi_\sigma,
\end{align}
where $T_{(\tm)}\coloneqq g^{\rho\sigma}T_{(\tm)\rho\sigma}$ is the trace of
$T_{(m)\rho\sigma}$, and $\alpha(\varphi)$ and $\beta(\varphi)$ were defined
in Eq.~\eqref{eq:alphabeta}.
Taking the divergence of Eq.~\eqref{eq:eins},
employing the contracted Bianchi identity $\nabla_\rho G^{\rho\sigma}=0$,
and using the scalar field equation of motion \eqref{eq:scalar}, we obtain
\begin{equation}
\label{eq:cons}
\nabla_\rho T_{(\tm)}^{\rho\sigma}
=-\nabla_\rho T_{(\varphi)}^{\rho\sigma}
=-{\cal Q}\varphi^\sigma,
\end{equation}
and the coupling strength ${\cal Q}$ can be rewritten as
\begin{equation}
\label{step1}
{\cal Q}=
\Lambda B^2(\varphi)
\left(\nabla_\rho T_{(\tm)}^{\rho\sigma}\right) \varphi_\sigma
+{\cal Y},
\end{equation}
where we have introduced
\begin{align}
{\cal Y} &\coloneqq
\Lambda B^2(\varphi)
\left\{
\left[
\beta (\varphi)
-\alpha (\varphi)
\right]
T_{(\tm)}^{\rho\sigma}
\varphi_\rho\varphi_\sigma
+
T_{(\tm)}^{\rho\sigma} \varphi_{\rho\sigma}
\right\} \nonumber \\
&-\alpha(\varphi)
T_{(\tm)}.
\end{align}
Multiplying Eq.~\eqref{eq:cons} by $\varphi_\sigma$
and solving it with respect to $\left(\nabla_\rho T_{(\tm)}^{\rho\sigma} \right)\varphi_\sigma$, we obtain
\begin{equation}
\chi
\left(
\nabla_\rho T^{\rho\sigma}_{(\tm)}
\right)\varphi_\sigma =2X{\cal Y}.
\end{equation}
Then, substituting it in Eq.~\eqref{step1}, using $\cal{Q}={\cal{Y}}/{\chi}$,
and finally eliminating $\cal{Q}$ from Eq.~\eqref{eq:scalar}, we obtain the
reduced scalar field equation of motion
\begin{align}
\label{eq:scalar_eq}
&P_X\Box\varphi+ P_\varphi
-P_{XX}\varphi^{\rho}\varphi^\sigma\varphi_{\rho\sigma}-2XP_{X\varphi}
=
\frac{\kappa}{\chi(X,\varphi)} \times \nonumber \\
&\left\{
\Lambda B^2(\varphi)
\left[
\left(
 \beta(\varphi)
-\alpha(\varphi)
\right)
T^{\rho\sigma}_{(\tm)}\varphi_\rho \varphi_\sigma
+
T^{\rho\sigma}_{(\tm)}\varphi_{\rho\sigma}
\right] \right. \nonumber \\
&\left. -\alpha (\varphi)
T_{(\tm)}
\right\}.
\end{align}

\section{The equations of stellar structure}
\label{sec:structure}

\subsection{Equations of motion}

In this section, we consider a static and spherically symmetric spacetime
with line element
\begin{align}
\label{eq:sss}
ds^2 & =g_{\mu\nu}dx^\mu dx^\nu \nonumber \\
&=-e^{\nu(r)}c^2 dt^2 +e^{\lambda(r)} dr^2+ r^2\gamma_{ij}d\theta^i d\theta^j,
\end{align}
where $\nu(r)$ and $\lambda(r)$ are functions of the radial coordinate $r$ only,
$\gamma_{ij}$ is the metric of the unit 2-sphere,
and the coordinates $\theta^i$ ($i=1,2$) run over the directions of the unit 2-sphere,
such that $\gamma_{ij}d\theta^i d\theta^j=d\theta^2+\sin^2\theta d\phi^2$.
We also assume by symmetry that the scalar field is only a function of $r$,
$\varphi=\varphi(r)$. Hence the coupling functions $A(\varphi)$ and $B(\varphi)$
are also only functions of $r$ through $\varphi(r)$.

We assume that in the Jordan frame only diagonal components of
the energy-momentum tensor of matter are nonvanishing
\begin{equation}
{\tilde T}_{(\tm)}{}^t{}_t=-\tilde \rho c^2,
\quad
{\tilde T}_{(\tm)}{}^r{}_r={\tilde p}_r,
\quad
{\tilde T}_{(\tm)}{}^i{}_j
={\tilde p}_t \delta^i{}_j,
\end{equation}
where ${\tilde \rho}$, $\tilde p_r$ and ${\tilde p}_t$
are respectively the energy density, radial and tangential pressures of
an anisotropic fluid in the Jordan frame~\cite{BowersLiang:1974}.
Using Eq.~\eqref{stu}, they are related to the components of the
energy-momentum tensor of matter in the Einstein frame, which are represented by
the quantities without a {\it tilde}, by
\begin{equation}
\rho = A^4(\varphi)
\sqrt{\chi}
{\tilde \rho},
\quad
p_r=
\frac{A^{4}(\varphi)}{\sqrt{\chi} }
{\tilde p}_r,
\quad
p_t
= A^{4}(\varphi)\sqrt{\chi}
{\tilde p}_t,
\label{relation}
\end{equation}
where in the background given by Eq.~\eqref{eq:sss}, the
quantity $\chi$ defined in Eq.~\eqref{eq:chi}
reduces to
\begin{equation}
\chi= 1+e^{-\lambda} \Lambda B^2(\varphi) (\varphi')^2.
\end{equation}
We note that even if the fluid in the Jordan frame
has an isotropic pressure,
${\tilde p}_r={\tilde p}_t$,
it is transformed into an anisotropic one in the Einstein frame
i.e. ${p}_r\neq {p}_t$ in the presence of disformal coupling $\chi\neq 1$.

The $(t,t)$, $(r,r)$ and the trace of $(i,j)$ components of the Einstein
equations~\eqref{eq:einstein_cov} are given by
\begin{align}
\label{eq:einst1}
&\frac{1}{r^2}
\left[
1-e^{-\lambda} (1-r\lambda')
\right]
=
- P + A^4(\vp) \sqrt{\chi} \kappa {\tilde \rho}c^2 , \\
&\frac{e^\lambda}{r^2}
\left[
1-e^{-\lambda} (1+r\nu')
\right]
=
-(\varphi')^2P_X -
e^{\lambda} \left[ P +\frac{A^4(\vp)}
 {\sqrt{\chi}}
 (\kappa {\tilde p}_r) \right]
 \nonumber \\
\label{eq:einst2}
\\
\label{eq:einst3}
&\frac{1}{2}
\left[
\nu''+\left(\frac{\nu'}{2}+\frac{1}{r}\right)(\nu'-\lambda')
\right]
=
e^\lambda \left[ P  + A^4(\vp)\sqrt{\chi}(\kappa {\tilde p}_t)\right].
\end{align}
\begin{widetext}
On the other hand,
the scalar field equation of motion \eqref{eq:scalar_eq} reduces to
\begin{eqnarray}
\label{eq:scalar_r1}
&&
\chi
\left\{
P_X e^{-\lambda}
\left[
\varphi''
+\left(\frac{\nu'}{2}-\frac{\lambda'}{2}+\frac{2}{r}\right)\varphi'
\right]
+P_\varphi
-P_{XX}
e^{-2\lambda} (\varphi')^2
\left(\varphi'' -\frac{\lambda'}{2}\varphi'\right)
+e^{-\lambda} (\varphi')^2 P_{X\varphi}
\right\}
\nonumber\\
&=&\kappa
\frac{A^{4}(\vp)}{\varphi'}
\left\{
\frac{{\tilde p}_r}
{\sqrt{\chi}}
\left[
-\alpha(\varphi) \varphi'
+\Lambda B^2(\vp) e^{-\lambda}
\varphi'
\left(\varphi''
+\left(
 \beta (\varphi)\varphi'
-\alpha(\varphi) \varphi'
-\frac{\lambda'}{2}
\right)
\varphi' \right)
\right] \right.
\nonumber\\
&-&\left.
\sqrt{\chi}
\left[
\alpha(\varphi)\varphi'
\left(-{\tilde \rho}c^2 +2{\tilde p}_t\right)
+\Lambda B^2(\vp) e^{-\lambda}
\left(
\frac{\nu'}{2}{\tilde\rho} c^2
-\frac{2}{r}{\tilde p}_t\right)
(\varphi')^2
\right]
\right\}.
\end{eqnarray}
\end{widetext}

The nontrivial radial component of the energy-momentum conservation law in the
Einstein frame~\eqref{eq:cons} gives us
\begin{align}
\frac{d{\tilde p}_r}{dr} =
-\left[\frac{\nu'}{2}
+\alpha (\varphi)\varphi'
\right]
\left({\tilde \rho}c^2 +{\tilde p}_r\right)
- 2\left[\frac{1}{r}
+\alpha (\varphi) \varphi'
\right]
{\tilde \sigma}, \nonumber \\
\label{eq:dpdr}
\end{align}
where we have defined ${\tilde \sigma}\coloneqq {\tilde p}_r-{\tilde p}_t$,
which measures the degree of anisotropy of the fluid~\cite{BowersLiang:1974}.
The same result can be obtained from the conservation law
in the Jordan frame ${\tilde \nabla}_\rho {\tilde T}_{(\tm)}^{\rho r}=0$, where
$\tilde \nabla_\rho$ represents the covariant derivative associated with the
Jordan frame metric ${\tilde g}_{\mu\nu}$. The conservation
law~\eqref{eq:dpdr} depends implicitly on $B(\varphi)$ and its derivative
through $\nu'$ [cf. Eq.~\eqref{eq:einst2}].

\subsection{The reduced equations of motion}

We then reduce the set of equations \eqref{eq:einst1}-\eqref{eq:einst3},
\eqref{eq:scalar_r1} and \eqref{eq:dpdr}
into a form more convenient for a numerical integration.
We introduce the mass function $\mu(r)$ through
\begin{equation}
e^{-\lambda(r)} := 1-\frac{2\mu (r)}{r},
\end{equation}
and replace all $\lambda(r)$ dependence with $\mu(r)$.
We also introduce the first-order derivative of the scalar field $\psi(r)$, i.e.
\begin{equation}
\label{eq0}
\psi \coloneqq \frac{d\varphi}{dr}.
\end{equation}
We can write the kinetic energy as
\begin{equation}
X= - \frac{r-2\mu}{2r}\psi^2
\end{equation}
and $\chi$ can then be expressed as
\begin{equation}
\chi = 1+\frac{r-2\mu}{r} \Lambda B^2(\varphi) \psi^2.
\end{equation}

The $(t,t)$ component of the Einstein equations
[cf. Eq~\eqref{eq:einst1}] determines the gradient of $\mu$
\begin{equation}
\label{eq1}
\frac{d\mu}{dr}
=\frac{r^2}{2}
\left[
A^4(\vp) \sqrt{\chi} \kappa {\tilde \rho}c^2 -P
\right].
\end{equation}
Similarly, the $(r,r)$ component of the Einstein equations \eqref{eq:einst2} reduces to
\begin{align}
\frac{d\nu}{dr}
&= \frac{2\mu}{r(r-2\mu)} \nonumber \\
&+r
\left\{
\psi^2 P_X
+\frac{r}{r-2\mu}
\left[
P
+\frac{A^4(\varphi)}{\sqrt{\chi}}
 (\kappa {\tilde p}_r)
\right]
\right\}.
\label{eq2}
\end{align}
\begin{widetext}
The conservation law~\eqref{eq:dpdr} combined with Eq.~\eqref{eq2} leads to
\begin{eqnarray}
\label{eq3}
\frac{d{\tilde p}_r}{dr}
&=&
-\left\{
\alpha(\varphi) \psi
+\frac{\mu}{r(r-2\mu)}
+\frac{r}{2}
\left[
\psi^2 P_X
+\frac{r}{r-2\mu}
\left(
P
+\frac{A^4(\vp)}{\sqrt{\chi}}
 (\kappa {\tilde p}_r)
\right)
\right]
\right\}
\left({\tilde \rho}c^2+{\tilde p}_r\right)
-2\left[\frac{1}{r}
+\alpha(\varphi)\psi
\right]
{\tilde \sigma}.
\end{eqnarray}
Finally, the scalar field equation of motion \eqref{eq:scalar_r1} reduces to
\begin{eqnarray}
\label{eq:scalar_r2}
&&
\left[
\chi
\left(P_X-e^{-\lambda}\psi^2P_{XX}\right)
-\kappa \Lambda A^4(\vp) B^2(\vp)
 \frac{{\tilde p}_r}{\sqrt{\chi}}
\right]
\psi'
+
\left\{
\chi
\left[
\left(\frac{\nu'}{2}-\frac{\lambda'}{2}+\frac{2}{r}\right)P_X
+\frac{\lambda'}{2}e^{-\lambda}\psi^2P_{XX}
+\psi P_{X\varphi}
\right] \right.
\nonumber\\
&-& \left.
\kappa \Lambda A^4(\vp) B^2(\vp)
\left[
\sqrt{\chi}
\left(
-\frac{\nu'}{2}{\tilde\rho} c^2
+\frac{2}{r}{\tilde p}_t
\right)
+\frac{{\tilde p}_r}{\sqrt{\chi}}
\left(
\beta(\varphi) \psi
-\alpha(\varphi)\psi
-\frac{\lambda'}{2}
\right)
\right]
\right\}
\psi
\nonumber\\
&=&
-e^{\lambda} \chi P_\varphi
+ \kappa A^4(\varphi) \alpha (\varphi)
e^{\lambda}
\left[
-\frac{{\tilde p}_r}{\sqrt{\chi}}
+\sqrt{\chi}
({\tilde\rho}c^2-2{\tilde p}_t)
\right].
\end{eqnarray}
Eliminating $\lambda'$ and $\nu'$ from Eq.~\eqref{eq:scalar_r2}, and using
Eqs.~\eqref{eq:einst1}-\eqref{eq:einst2},
the scalar field equation of motion~\eqref{eq:scalar_r2} can be rewritten as
\begin{equation}
\label{eq_s}
C_2 \frac{d\psi}{dr}
=-C_1\psi
+\frac{r}{r-2\mu}
\left\{
-\chi
P_\varphi+ \kappa A^4(\varphi)\alpha(\varphi)
\left[
-\frac{{\tilde p}_r}{\sqrt{\chi}}
-\sqrt{\chi}
(-{\tilde\rho} c^2 +2{\tilde p}_t)
\right]
\right\},
\end{equation}
where we introduced
\begin{align}
C_2
&=
\chi \left[P_X -\left(1-\frac{2\mu}{r}\right) \psi^2 P_{XX}\right]
-\kappa \Lambda  A^4(\varphi) B^2(\varphi)
 \frac{{\tilde p}_r}{\sqrt{\chi}},
\nonumber\\
C_1&=
\chi
\left\{
P_X
\left[
\frac{2(r-\mu)}{r(r-2\mu)}
+\frac{r}{2} \psi^2P_X
+\frac{r^2}{r-2\mu}
\left(
P
-
\frac{\kappa}{2} A^4(\varphi)
\left(
\sqrt{\chi} {\tilde \rho}c^2-\frac{{\tilde p}_r}{\sqrt{\chi}}
\right)
\right)
\right]
\right.
\nonumber\\
&\left. +\frac{1}{2}
\left[
-\frac{2\mu}{r^2}
+r
\left(
-P
+A^4(\varphi)\sqrt{\chi}(\kappa {\tilde \rho}c^2)
\right)
\right]
\psi^2 P_{XX}
+\psi P_{X\varphi}\right\}
\nonumber\\
&-
\kappa \Lambda A^4(\varphi) B^2(\varphi)
\left\{
-\frac{1}{r-2\mu}
\left(\frac{\mu}{r}
+\frac{r^2 P}{2}
\right)
\left(
\sqrt{\chi}{\tilde\rho} c^2
-\frac{{\tilde p}_r}{\sqrt{\chi}}
\right)
-\frac{{\tilde \rho}c^2 \sqrt{\chi}}{2} \psi^2 P_X r \right.
\nonumber \\
&\left. - \frac{\kappa r^2}{r-2\mu}({\tilde\rho} c^2 {\tilde p}_r) A^4(\varphi)
+\frac{2\sqrt{\chi}}{r}{\tilde p}_t
+\frac{\psi}{\sqrt{\chi}}
\left(\beta (\varphi)
-\alpha(\varphi)
\right)
{\tilde p}_r
\right\}.
\end{align}
\end{widetext}
The set of Eqs.~\eqref{eq0}, \eqref{eq1}, \eqref{eq2}, \eqref{eq3} and \eqref{eq_s}
together with a given EOS
\begin{eqnarray}
{\tilde p}_r={\tilde p}_r(\tilde\rho),
\quad
{\tilde p}_t={\tilde p}_t(\tilde\rho),
\end{eqnarray}
form a closed system of equations to analyze the structure of relativistic stars in the
scalar-tensor theory~\eqref{eq:e_ac}.

\subsection{Slowly rotating stars}

In this subsection, we extend our calculation to the case of
slowly rotating stars. Once the set of the equations of motion for a static and
spherically symmetric star is given, it is simple to take first-order
corrections due to rotation into consideration using the
Hartle-Thorne scheme~\cite{Hartle:1967he,Hartle:1968si}.
At first order in the Hartle-Thorne perturbative expansion,
we derive our results in a manner as general as possible, similarly
to the previous section.

In the Einstein frame, the line element including the first-order correction due
to rotation is given by
\begin{align}
ds^2 &= -e^{\nu(r)} c^2 dt^2 + e^{\lambda(r)} dr^2 + r^2 \left( d\theta^2
+ \sin^2\theta d\phi^2 \right)
 \nonumber \\
&+2\left(\omega-\Omega\right)
r^2 \sin^2\theta dt d\phi,
\end{align}
where $\omega(r)$ is a function of $r$, which is of the same order as the star's angular velocity $\Omega$. We can construct the Jordan frame line element using
Eqs.~(\ref{eq:disformal}) and~(\ref{eq:inv_disformal}). The construction of the
energy-momentum tensor for the anisotropic fluid in the Jordan frame
is similar to what was done before, except that now, the normalization of the four-velocity, demands that
\begin{subequations}
\begin{align}
\tilde{u}^{t} &= \left[ -\left(\tilde{g}_{tt} + 2\tilde \Omega\tilde{g}_{t\phi}
+ \tilde \Omega^2\tilde{g}_{\phi\phi}\right) \right]^{-1/2},
\\
\tilde{u}^{r} &= \tilde{u}^{\theta} = 0,
\quad
\tilde{u}^{\phi} =\tilde  \Omega \tilde{u}^{t},
\end{align}
\end{subequations}
where $\tilde \Omega$ is the star's angular velocity in the Jordan frame
[measured in the coordinates of $x^{\mu}=(t,r,\theta,\phi)$],
\begin{subequations}
\begin{align}
{\tilde g}_{tt}&=A^2 g_{tt},
\quad
{\tilde g}_{rr}=A^2 \left[g_{rr}+\Lambda B^2\left(\varphi'\right)^2 \right],
\\
{\tilde g}_{ij}&=A^2 g_{ij},\quad (i,j=\theta,\phi)
\\
{\tilde g}_{t\phi}&=A^2 g_{t\phi},
\end{align}
\end{subequations}
and we must expand all expressions, keeping only terms of order
${\cal O}({\Omega})$. As shown in the Appendix the star's angular velocity is
disformally invariant, $\tilde \Omega=\Omega$.
We also note that rotation can induce a dependence of the scalar field
on $\theta$, which appears however only at more than second order in rotation,
${\cal O} (\Omega^2)$ ~\cite{Pani:2014jra}. Thus in our case, the scalar field
configuration remains the same as in the nonrotating situation.

At the first order in rotation, the diagonal components of the Einstein equations
and the scalar field equation of motion remain the same as Eqs.
\eqref{eq0}, \eqref{eq1}, \eqref{eq2}, \eqref{eq3} and \eqref{eq_s}.
A new equation comes however from the $(t,\phi)$ component of
the Einstein equation:
\begin{align}
&\frac{d^2\omega}{dr^2}
 -
\left(\frac{4}{r}-\frac{\lambda'+\nu'}{2}\right)
\frac{d\omega}{dr} \nonumber \\
&+2\kappa A^4(\varphi)
r\sqrt{\chi} \frac{\left(\tilde{\rho} c^2 + \tilde{p}_{r} - \tilde{\sigma}\right)}{(r - 2\mu)}\,
\omega(r) = 0.
\label{eq:framedrag0}
\end{align}
By eliminating $\nu'$ and $\lambda'$ with the use of Eqs.~\eqref{eq1} and
\eqref{eq2}, we obtain the frame-dragging equation
\begin{align}
\frac{d^2\omega}{dr^2}
&+ \left[ \frac{1}{2} r P_{X} \psi^2
+ \frac{
\kappa r^2 A^4(\varphi)}{2\sqrt{\chi}(r-2\mu)} \left({\tilde{p}_r} + \chi\,\tilde{\rho}c^2 \right)
- \frac{4}{r}\right]
\frac{d \omega}{dr} \nonumber \\
&+
2\kappa A^4(\varphi)
r\sqrt{\chi} \frac{\left(\tilde{\rho}c^2 + \tilde{p}_{r} - \tilde{\sigma}\right)}{(r - 2\mu)}\,
\omega(r) = 0.
\label{eq:framedrag}
\end{align}
Equation~\eqref{eq:framedrag} can be solved together with
Eqs.~\eqref{eq0}, \eqref{eq1}, \eqref{eq2}, \eqref{eq3} and \eqref{eq_s}.
Together these equations fully describe a slowly rotating anisotropic relativistic
star in the theory described by the action~\eqref{eq:e_ac}.

\subsection{Particular limits}

The equations obtained in the previous section represent the
most general set of stellar structure equations for a broad class of
scalar-tensor theories with a single scalar degree of freedom with a
disformal coupling between the scalar field and a
spherically symmetric slowly rotating anisotropic fluid distribution.
Because of its generality, we can recover many particular cases
previously studied in the literature:

\begin{enumerate}
\item In the limit of the pure conformal coupling, $\Lambda\to 0$ (thus $\chi\to 1$),
we recover the case studied in Ref.~\cite{Silva:2014fca}.
\item If we additionally assume isotropic pressure
${\tilde p}_r={\tilde p}_t={\tilde p}$, we recover the standard equations
given in Refs.~\cite{Damour:1993hw,Damour:1996ke}.
\item If we assume a kinetic term of the form
$P(X,\varphi) = 2 X -V(\varphi)$, where $V(\varphi)$
is a mass term $m^2\varphi^2$, isotropic pressure
and purely conformal coupling  we recover the massive
scalar-tensor theory studied in Refs.~\cite{Ramazanoglu:2016kul,Yazadjiev:2016pcb}
and the asymmetron scenario proposed in Ref.~\cite{Chen:2015zmx} by
appropriately choosing $A(\vp)$.
\end{enumerate}

\section{Scalar-tensor theory with a canonical scalar field}
\label{sec:case}

\subsection{Stellar structure equations}

Now let us apply the general formulation developed in the previous
section to the canonical scalar field with the potential $V(\varphi)$,
i.e. $P=2 X-V(\varphi)$. The stellar structure equations~\eqref{eq0},
\eqref{eq1}, \eqref{eq2}, \eqref{eq3} and~\eqref{eq_s}
reduce to
%
\begin{subequations}
\label{set}
\begin{align}
\frac{d\mu}{dr}
&=\frac{r(r-2\mu)}{2} \psi^2
+\frac{r^2}{2}V(\varphi)
+A^4(\varphi)\sqrt{\chi} \left(\frac{\kappa}{2}{\tilde \rho} c^2 r^2\right),
\nonumber \\
\\
\frac{d\nu}{dr}
&= \frac{2\mu}{r(r-2\mu)}
+r\psi^2
-\frac{r^2}{r-2\mu}V(\varphi) \nonumber \\
&+\frac{r^2}{r-2\mu}
 \frac{A^4(\varphi)}{\sqrt{\chi}}
 (\kappa {\tilde p}_r),
\\
\frac{d{\tilde p}_r}{dr}
&=
-\left[
\alpha(\varphi)\psi
+\frac{\mu}{r(r-2\mu)}
+\frac{r}{2} \psi^2
-\frac{
r^2}{2(r-2\mu)}V(\varphi) \right. \nonumber \\
&\left.+\frac{r^2}{r-2\mu}
  \frac{A^4(\varphi)}{\sqrt{\chi}} \left(\frac{\kappa}{2}{\tilde p}_r\right)
\right]
\left({\tilde \rho}c^2 +{\tilde p}_r\right)
\nonumber \\
&-2\left(\frac{1}{r} +\alpha(\varphi) \psi \right)
{\tilde \sigma},
\\
\frac{d\varphi}{dr}&=\psi,
\\
\label{sca_cano}
C_2 \frac{d\psi}{dr}
&=-C_1\psi
+\frac{
r\chi V_\varphi (\varphi)}{r-2\mu}
+ \frac{\kappa r}{r-2\mu}
A^{4}(\varphi)\alpha(\varphi) \times \nonumber \\
&\left[
-\frac{{\tilde p}_r}{\sqrt{\chi}}
+\sqrt{\chi}
({\tilde\rho}c^2 -2{\tilde p}_r)
+2\sqrt{\chi}
{\tilde\sigma}
\right],
\end{align}
\end{subequations}
where
\begin{align}
C_1 &= \frac{2 \chi}{r - 2 \mu}
\left[ \frac{2(r - \mu)}{r} - r^2 V(\vp) - \frac{\kappa}{2} A^4(\vp)r^2 \right. \nonumber \\
&\times \left. \left( \sqrt{\chi} \tilde{\rho} c^2 - \frac{\tilde{p}_r}{\sqrt{\chi}}   \right) \right] - \kappa \Lambda A^4(\vp) B^2(\vp) \nonumber \\
&\times \left[ - \frac{\mu}{r(r - 2\mu)}\left( \sqrt{\chi} \tilde{\rho} c^2 - \frac{\tilde{p}_r}{\sqrt{\chi}}   \right) \right. \nonumber \\
&\left. - \frac{r\psi^2}{2}\left( \sqrt{\chi} \tilde{\rho} c^2 + \frac{\tilde{p}_r}{\sqrt{\chi}} \right) + \frac{r^2 V(\vp)}{2(r-2\mu)} \right.
\nonumber \\
&\times \left. \left( \sqrt{\chi} \tilde{\rho} c^2 - \frac{\tilde{p}_r}{\sqrt{\chi}} \right) - \kappa A^4(\vp)\frac{ r^2}{r - 2\mu} \tilde{p}_r \tilde{\rho} c^2
\right. \nonumber \\
&+ \frac{2 \sqrt{\chi}}{r}(\tilde{p}_r - \tilde{\sigma})
+\left. \frac{\psi}{\sqrt{\chi}} (\beta(\vp) - \alpha(\vp))\tilde{p}_r \right]
\label{eq:c1}
\end{align}
and
\begin{equation}
C_2
= 2\chi -\kappa \Lambda A(\varphi)^4B(\varphi)^2 \frac{{\tilde p}_r}{\sqrt{\chi}}.
\label{eq:c2}
\end{equation}

In the case of a slowly rotating star,
the frame-dragging equation~\eqref{eq:framedrag} becomes
\begin{align}
&\frac{d^2\omega}{dr^2}
- \left[
 r \psi^2
+ \frac{
\kappa r^2 A^4(\varphi)}{2(r-2\mu)}
\left(\frac{\tilde{\rho}c^2}{\sqrt{\chi}} + \sqrt{\chi} \tilde{p}_r \right)
- \frac{4}{r}\right]
\frac{d\omega}{dr} \nonumber \\
&- 2\kappa A^4(\varphi) r\sqrt{\chi} \frac{\left(\tilde{\rho}c^2 + \tilde{p}_{r}
-\tilde{\sigma}\right)}{(r - 2\mu)}\,
\omega(r) = 0.
\label{eq:framedrag2}
\end{align}

Through the Einstein equation \eqref{eq:einst2}, we find that if
$\Lambda>0$ the second term of $C_2$ in Eq.~\eqref{eq:c2}
is of order ${\cal O}\left({\Lambda B^2}/{r^2}\right)$,
from which we
can estimate the radius within which
the contributions of disformal coupling to the gradient terms
become comparable to the standard ones in the scalar-tensor theory as
$R_{\tcd}:= \sqrt{\Lambda} B(\varphi)$.
If $R_\tcd>R$, where $r=R$ is the star's radius,
the contributions of disformal coupling to the gradient terms
become important throughout the star,
while if $R_\tcd<R$ they
could be important only in a portion of the star's interior $r<R_\tcd$.
When $B\to 1$,
$R_\tcd \approx \sqrt{\Lambda}$ and
therefore $\sqrt{\Lambda}$ characterizes the length scale
for which the disformal coupling effects become apparent.
As the radius of a typical NS is about $10$ km, the effects of disformal coupling
of the star become apparent when $\Lambda> {\cal O}(100\,{\rm km}^2)$.

We note that in the presence of the disformal coupling,
when integrating the scalar field equation~\eqref{eq_s},
the coefficient $C_2$ in the $d\psi/dr$ equation may vanish at
some $r=R_\ast$, i.e. $C_2(R_\ast)=0$.
This could happen when both $\Lambda>0$ and the pressure at the center of the
star is large enough such that $C_2<0$ in the vicinity of $r=0$.
In such a case, as we integrate the equations outwards, since the radial pressure ${\tilde p}_r$ decreases and vanishes at the surface of the star, there must be a
point $R_{\ast}$ where $C_2$ vanishes. This point represents a singularity of our
equations and a regular stellar model cannot be constructed.
The nonexistence of a regular relativistic star for a large positive
$\Lambda$ is one of the most important consequences due to the disformal coupling.
The appearance of the singularity is due to the fact that
the gradient term in the scalar field equation of motion~\eqref{sca_cano}
picks a wrong sign (i.e., negative speed of sound) and is an illustration of the
gradient instability pointed out in Refs.~\cite{Koivisto:2012za,Berezhiani:2013dw,Bettoni:2015wta}.

\subsection{Interior solutions}
\label{sec:interior}

From this section onwards, we focus on the case of isotropic pressure
${\tilde p}={\tilde p}_r={\tilde p}_t$.
We then derive the boundary conditions at the center of the star, $r=0$,
which have to be specified when integrating Eqs. \eqref{set}
and \eqref{eq:framedrag2}.
We assume that at $r=0$, ${\tilde \rho}(0)={\tilde\rho}_\tc$.
The remaining metric and matter variables can be expanded as
\begin{widetext}
\begin{subequations}
\label{centera}
\begin{align}
\mu(r)&=\frac{1}{6}
\left[\kappa {\tilde \rho}_\tc c^2 A^4(\varphi_\tc)+ V(\varphi_\tc)
\right] r^3+ {\cal O}(r^5),
\\
\nu(r)&=
\frac{1}{6}
\left[\kappa\left( {\tilde \rho}_\tc c^2 +3 {\tilde p}_\tc\right)
A^4(\varphi_\tc)
-2V(\varphi_\tc)
\right] r^2
+{\cal O}(r^4),
\\
\label{vp1}
\varphi (r)&=\varphi_\tc
+\frac{ \kappa  A^4(\varphi_\tc)\alpha(\varphi_\tc)
\left({\tilde \rho}_\tc c^2-3{\tilde p}_\tc\right)+V_\varphi(\varphi_\tc)}
{12\left[2-{\kappa}\Lambda {\tilde p}_\tc A^4(\varphi_\tc) B^2(\varphi_\tc)\right]}
 r^2
+{\cal O}(r^4),
\\
\label{p1}
{\tilde p} (r)
&=
{\tilde p}_\tc
-\frac{1}{12} \left(
{\tilde \rho}_\tc c^2 +{\tilde p}_\tc\right)
\left\{
\kappa A^4(\varphi_\tc)
\left[
{\tilde \rho}_\tc c^2 + 3{\tilde p}_\tc
+\alpha (\varphi_\tc)^2
\frac{{\tilde \rho}_\tc c^2 -3{\tilde p}_\tc}
        {1-\frac{\kappa}{2}\Lambda {\tilde p}_\tc A^4(\varphi_\tc) B^2(\varphi_\tc)}
\right] \right.
\nonumber \\
&\left. -2 V(\varphi_\tc)
\left[
1
-\frac{\alpha (\varphi_\tc) V_\varphi(\varphi_\tc)}{2V(\varphi_\tc)}
\frac{1}{1-\frac{\kappa}{2}\Lambda {\tilde p}_\tc A^4(\varphi_\tc) B^2(\varphi_\tc)}
\right]
\right\}
r^2
+{\cal O}(r^4),
\end{align}
\end{subequations}
\end{widetext}
where ${\tilde p}_\tc$ is fixed by ${\tilde \rho}_\tc$ through the EOS, i.e.
${\tilde p}_\tc={\tilde p}(\tilde\rho_\tc)$. The central value of the scalar field
$\varphi_\tc$ is fixed by demanding that outside the star the scalar field approaches
a given cosmological value $\varphi_0$ as $r\to \infty$, which is consistent with
observational constraints. We will come back to this in Sec.~\ref{sec:exterior}.

As a well-behaved stellar model requires ${\tilde p}''(0)<0$, we impose
\begin{align}
&\kappa A^4(\varphi_\tc)
\left[{\tilde \rho}_\tc c^2+3{\tilde p}_\tc
+\alpha (\varphi_\tc)^2
\frac{{\tilde \rho}_\tc c^2-3{\tilde p}_\tc}
        {1-\frac{\kappa}{2}\Lambda {\tilde p}_\tc A^4(\varphi_\tc) B^2(\varphi_\tc)}
\right]
\nonumber\\
&-2V(\varphi_\tc)
\left[
1
-\frac{\alpha (\varphi_\tc) V_\varphi (\varphi_\tc)}{2V(\varphi_\tc)}
\frac{1}{1-\frac{\kappa}{2}\Lambda {\tilde p}_\tc A^4(\varphi_\tc) B^2(\varphi_\tc)}
\right] \nonumber \\
&>0.
\end{align}
For a large positive disformal coupling parameter $\Lambda>0$
and a large pressure at the center $\tilde p_\tc$
such that $\left|1-\frac{\kappa \Lambda}{2} {\tilde p}_\tc A^{4}(\varphi_\tc) B^{2}(\varphi_\tc)\right| \ll 1$,
the $r^2$ terms of the scalar field and pressure diverge and
the Taylor series solution~\eqref{centera} breaks down.
Such a property is a direct consequence of the appearance of the singularity inside the star which was mentioned in the previous subsection.
Assuming that $A(\varphi_\tc)\approx 1$ and $B(\varphi_\tc)\approx 1$,
the maximal positive value of $\Lambda_{\rm max}$ can be roughly estimated as
\begin{align}
\Lambda_{\rm max}
\approx
\frac{2}{\kappa{\tilde p}_\tc}
=\frac{c^4}{4\pi G{\tilde p}_\tc}
\approx 10^2\, {\rm km}^2,
\end{align}
for $\tilde p_\tc=10^{36}$ dyne/cm${}^2$,
which agrees with the numerical analysis done in
Sec.~\ref{sec:results}.
On the other hand, for a large negative value of the disformal coupling $\Lambda<0$,
no singularity appears, from Eq.~\eqref{vp1}
the $r^2$ correction to the scalar field amplitude is suppressed,
and $\varphi(r)\to \varphi_\tc$ everywhere inside the star.
This indicates that $A(\varphi_\tc)\approx {\rm constant}$, and
for a vanishing potential $V(\varphi)=0$ the stellar configuration approaches
that in GR.

In the case of slowly rotating stars, the boundary condition
for $\omega$ near the origin reads
\begin{equation}
\omega
\label{bc_omega}
=\omega_\tc
\left[
1
+\frac{\kappa}{5} A^{4}(\varphi_\tc)
\left(
{\tilde \rho}_\tc c^2 +{\tilde p}_\tc
\right)r^2
\right]+{\cal O}(r^4).
\end{equation}

\vspace{0.1cm}
\subsubsection{Stellar models in purely disformal theories}
\label{sec:purelydis}
It is interesting to analyze the stellar structure equations in the purely disformal
coupling limit, when $A(\varphi)=1$. In this case we find that the expansions
near the origin are
\begin{align}
\label{center2}
\mu(r)&=\frac{1}{6}
\left[\kappa {\tilde \rho}_\tc c^2 +
V(\varphi_\tc)\right] r^3+ {\cal O}(r^5),
\nonumber\\
\nu(r)&=
\frac{1}{6}
\left[\kappa\left( {\tilde \rho}_\tc c^2 +3 {\tilde p}_\tc\right)
-2 V(\varphi_\tc)
\right] r^2
+{\cal O}(r^4),
\nonumber\\
\varphi (r)&=\varphi_\tc
+\frac{ V_\varphi(\varphi_\tc)}
{12\left[1-\frac{\kappa}{2}\Lambda {\tilde p}_\tc  B^2 (\varphi_\tc)\right]}
 r^2
+{\cal O}(r^4),
\nonumber\\
{\tilde p} (r)
&=
{\tilde p}_\tc
-\frac{1}{12} \left({\tilde p}_\tc+{\tilde \rho}_\tc c^2\right)
\left[
\kappa\left({\tilde \rho}_\tc c^2+3{\tilde p}_\tc\right)
-2V(\varphi_\tc)
\right]r^2 \nonumber \\
&+{\cal O}(r^4),
\end{align}
Thus for $V(\varphi)=0$, $\varphi=\varphi_\tc$ everywhere,
and the disformal coupling term does not modify the stellar structure with respect
to GR. Only with a nontrivial potential $V(\varphi)$, the disformal coupling can
modify the profile of the scalar field inside the NS. It was argued
in Ref.~\cite{Sakstein:2014isa} that for a simple mass term potential
$V_{\vp} \sim m^2 \vp$, where $m$ is the mass of the scalar field,
disformal contributions can be neglected and the NS solution is the same as
in GR.

\subsubsection{Metric functions in the Jordan frame}
Finally, we mention the behaviors of the metric functions in the Jordan frame.
In the Appendix we derive the relationship of the physical
quantities defined in the two frames. The boundary conditions~\eqref{centera}
indicate that in the singular stellar solution of the Einstein frame
the metric functions $\mu$ and $\nu$ remain regular. Using Eqs.~\eqref{f1}
and~\eqref{bmr}, the metric functions in the Jordan frame behave as
\begin{widetext}
\begin{align}
{\bar \nu}(r)
&=
\ln A(\varphi_\tc)^2
+\frac{1}{6}
\left[
\left({\tilde \rho}_\tc c^2+3{\tilde p}_\tc\right)-2V(\varphi_\tc)
+\alpha(\varphi_\tc)
\frac{\kappa A^{4}(\varphi_\tc)\alpha \left(\tilde \rho_\tc c^2-3 {\tilde p}_\tc\right)+V'(\varphi_\tc)}
      {1-\frac{\kappa \Lambda}{2} A^{4}(\varphi_\tc) B^{2}(\varphi_\tc) {\tilde p}_\tc}
\right]
r^2
+O(r^4),
\\
{\bar \mu}(r)
&=
\frac{A(\varphi_\tc)}{18}
\left[
3\left(A^4(\varphi_\tc) {\tilde \rho}c^2 +V(\varphi_\tc)\right)
+3\alpha(\varphi_\tc)
\frac{\kappa A^4(\varphi_\tc) \alpha \left(\tilde \rho_\tc c^2-3 {\tilde p}_\tc\right)+V'(\varphi_\tc)}
{1-\frac{\kappa \Lambda}{2} A^4(\varphi_\tc) B^2(\varphi_\tc) {\tilde p}_\tc}
\right.
\nonumber\\
&+\left. \Lambda B^2(\varphi_\tc)
\frac{\left(\kappa A^{4}(\varphi_\tc) \alpha\left(\tilde \rho_\tc c^2-3 {\tilde p}_\tc\right)
+V'(\varphi_\tc)\right)^2}
{4\left(1-\frac{\kappa \Lambda}{2} A^{4}(\varphi_\tc) B^{2}(\varphi_\tc) {\tilde p}_\tc\right)^2}
\right]
r^3
+O(r^5).
\end{align}
\end{widetext}
Therefore,
for $\left|1-\frac{\kappa \Lambda}{2} A^{4}(\varphi_\tc) B^{2}(\varphi_\tc) {\tilde p}_\tc\right|\ll 1$,
the Taylor series solutions for ${\bar \mu}(r)$ and ${\bar \nu}(r)$
break down, which indicates
that the metric functions in the Jordan frame $\bar \mu$ and $\bar \nu$ diverge
at some finite radius and a curvature singularity appears there.

\subsection{Exterior solution}
\label{sec:exterior}

In the vacuum region outside the star $r>R$,
the fluid variables ${\tilde \rho}$, ${\tilde p}_r$ and ${\tilde p}_t$ vanish.
The exterior solution should be the vacuum solution of GR
coupled to the massless canonical scalar field.
The following exact solution can be obtained~\cite{Damour:1993hw,Damour:1995kt}
\begin{align}
\label{eq:rho}
ds^2&=-e^{\nu(\rho)}c^2dt^2+ e^{-\nu(\rho)}
\nonumber \\
&\times \left[d\rho^2
+ \left(\rho^2-\frac{2Gs}
{c^2}\rho\right)\gamma_{ij}d\theta^i d\theta^j\right],
\\
\nu(\rho)
&=\nu_0
+\ln \left(1-\frac{2Gs}
{c^2\rho}\right)^{\frac{M}{s}},
\\
\varphi(\rho)
&=
\varphi_0
-\frac{Q}{2M}
\ln\left(1-\frac{2Gs}
{c^2 \rho}\right)^{\frac{M}{s}},
\end{align}
where $\nu_0$ represents the freedom of the rescaling of the time
coordinate, $\varphi_0$ is the cosmological value of the scalar field at
$r\to \infty$, 
$M$ and $Q$ are the integration constants and $s\coloneqq \sqrt{M^2+Q^2}$.
The metric \eqref{eq:rho} can be rewritten in terms of the Schwarzschild-like
coordinate $r$ by the transformations
\begin{align}
\label{rho3}\
r(\rho)
&=
\rho
\left(1-\frac{2Gs}
{c^2 \rho}\right)^{\frac{s-M}{2s}},
\\
\mu(\rho)
&=
M\left[
1
-\frac{G\left(s-M\right)^2}
{2M \rho c^2\left(1-\frac{2Gs}{c^2 \rho}\right)}
\right]
\left(1-\frac{2Gs}
{c^2 \rho}\right)^{\frac{s-M}{2s}}.
\end{align}
As $r\to\infty$, the solution \eqref{eq:rho} behaves as
\begin{subequations}
\label{def:massandam}
\begin{align}
\mu(r)&=
\frac{GM}{c^2}
-\frac{G^2Q^2}{2c^4 r}
+{\cal O}\left(\frac{1}{r^2}\right),
\\
\nu(r)&=
\nu_0
-\frac{2GM}{c^2r}
+{\cal O}\left(\frac{1}{r^2}\right),
\\
\varphi(r)
&=
\varphi_0
+\frac{GQ}{c^2r}
+{\cal O}\left(\frac{1}{r^2}\right).
\end{align}
\end{subequations}
Thus the integration constants $M$ and $Q$ correspond to the
Arnowitt-Deser-Misner (ADM) mass and the scalar charge in the
Einstein frame, respectively. For later convenience we also define
the fractional binding energy
\begin{equation}
{\cal E}_\tb  \coloneqq \frac{M_\tb}{M} - 1,
\label{eq:binding}
\end{equation}
which is positive for bound (but not
necessarily stable) configurations.
We note that for the vanishing scalar field at asymptotic infinity
the ADM mass is disformally invariant, $\bar M=M$
[see Eq.~\eqref{adm_equiv}].

In the slowly rotating case, the integration of Eq.~\eqref{eq:framedrag0} in
vacuum $\tilde{\rho}={\tilde p}_t=0$ gives
\begin{equation}
\label{eq:op}
\omega'= \frac{6G}{c^2 r^4} e^{\frac{\lambda+\nu}{2}}J,
\end{equation}
where $J$ is the integration constant.
In the vacuum case, we can find the exact exterior solution at the first order
in rotation~\cite{Damour:1996ke}. Expanding it in the vicinity of $r\to \infty$
gives
\begin{equation}
\omega=\Omega-\frac{2GJ}{c^2 r^3}+O\left(\frac{1}{r^5}\right).
\end{equation}
Thus $J$ corresponds to the angular momentum in the exterior spacetime.

\subsection{Matching}

At the surface of the star, the interior solution is matched to the exterior
solution~\eqref{eq:rho}. Then the cosmological value of the scalar field
$\varphi_0$, the ADM mass $M$ and the scalar charge $Q$ are evaluated as
\begin{subequations}
\begin{align}
\varphi_0
&=\varphi_\ts
+
\ln
\left(\frac{x_1+x_2} {x_1-x_2}
\right)^{\frac{\psi_\ts}{x_2}},
\\
M
&=
\frac{c^2 R^2
\nu_\ts'}{2G}
\left(1-\frac{2\mu_\ts}{R}
\right)^{\frac{1}{2}}
\left(
\frac{x_1+x_2}
      {x_1-x_2}
\right)^{-\frac{\nu'_\ts}{2x_2}},
\\
q & \coloneqq\frac{Q}{M}=-\frac{2\psi_\ts}{\nu_\ts'}.
\end{align}
\label{eq:surface_mat}
\end{subequations}
where we introduced $x_1 \coloneqq\nu_\ts' + {2}/R$ and
$x_2\coloneqq\sqrt{\nu_\ts'{}^2+4\psi_\ts^2}$. We also  defined
$\mu_\ts \coloneqq \mu(R)$ and $\nu_\ts \coloneqq\nu(R)$

In the case of a slowly rotating
star, the angular velocity and angular momentum of the star,
$\Omega$ and $J$, are evaluated as
\begin{align}
\Omega
&=
\omega_s
-\frac{3c^4 J}{4G^2 M^3 (3-\alpha(\varphi_\ts)^2)}
\left[
\frac{4}{x_1^2-x_2^2}
\left(\frac{x_1-x_2}
             {x_1+x_2}
\right)^{\frac{2\nu_\ts'}{x_2}}
\right. \nonumber \\
&\left. \times
\left(
\frac{3\nu_\ts'}{R}
+\frac{1}{R^2}
+3\nu_\ts'{}^2
-\psi_\ts^2
\right)
-1
\right],
\\
J&=\frac{c^2 R^4}
{6G} \sqrt{1-\frac{2\mu_\ts}{R}}
 e^{-\frac{\nu_\ts'}{2}}\omega_\ts'.
\end{align}
The moment of inertia can be obtained by
\begin{equation}
I \coloneqq \frac{J}{\Omega},
\end{equation}
or equivalently by integrating Eq.~\eqref{eq:framedrag0}, using Eqs.~\eqref{eq1} and
\eqref{eq2}
\begin{equation}
I
=\frac{8\pi}{3c^2}
\int_0^{R}
dr
A^{4}(\vp)\sqrt{\chi} \,r^4 e^{-\frac{\nu-\lambda}{2}}
\left({\tilde\rho}c^2 +{\tilde p}_t\right)
\left(\frac{\omega}{\Omega}\right).
\label{moi}
\end{equation}
We observe that this relation for the moment of inertia holds for any choice of
$P(X,\varphi)$, $A(\varphi)$ and $B(\varphi)$. In the purely conformal
theory we obtain the result of Ref.~\cite{Silva:2014fca}.

For a given EOS the equations of motion~\eqref{set} and~\eqref{eq:framedrag2}
are numerically integrated from $r=0$ up to the surface of the star
$r=R$, where the pressure vanishes $\tilde p(R)=0$. With the values of
various variables at the surface at hand, we can compute $\vp_0$, $M$, $q$
and $I$ using the matching conditions.

From the Einstein frame radius $R$, we can calculate the physical Jordan
frame radius $\tilde{R}$ through [cf. Eq.~\eqref{eq:disformal}]
\begin{equation}
\tilde{R} :=
\sqrt{
A^{2}(\vp_\ts)
\left[
 R^2
+ \Lambda B^{2}(\vp_\ts) \psi_\ts^2
\right]
}
\label{eq:radiusjordan}
\end{equation}
where we introduced $\vp_\ts \coloneqq \vp(R)$
and $\psi_\ts \coloneqq \psi(R)$. For a vanishing scalar
field we have $\tilde{R} = R$.

The total baryonic mass of the star $M_\tb$ can be obtained by integrating
\begin{equation}
M_\tb
= \int_{0}^{R}dr
A^3 (\vp) \sqrt{\chi}  \frac{4\pi {\tilde m}_\tb r^2}{\sqrt{1-\frac{2\mu}{r}}}{\tilde n}(r),
\end{equation}
where $\tilde{m}_\tb= 1.66 \times 10^{-24}$ g is the atomic mass unit
and ${\tilde n}$ is the baryonic number density.

In the Appendix we show that the physical quantities related to the rotation of
fluid and spacetime, namely $I$ and $J$ as well as $\omega$ and $\Omega$,
are invariant under the disformal transformation \eqref{eq:disformal}.

\section{A toy model of spontaneous scalarization with an incompressible fluid}
\label{sec:incompress}

Before carrying out the full numerical integrations of the stellar
structure equations it is illuminating to study under which conditions
scalarization can occur in our model. This can be accomplished by studying
a simple toy model where a scalar field lives on the background of
an incompressible fluid star.
The results obtained in this section will be validated
in Sec.~\ref{sec:results}.

Let us start by assuming that the star has a constant density
${\rho}$ (incompressible) and an isotropic pressure ${p}={p}_r={p}_t$. The
scalar field $\varphi$ is massless, and has a canonical kinetic term
and small amplitude, such that we can linearize the equations of motion. The
conformal and disformal coupling functions can be expanded as
\begin{eqnarray}
\label{def:ab}
A(\varphi)&=&1 +\frac{1}{2}\beta_1\varphi^2+{\cal O}\left(\varphi^3\right),
\nonumber\\
B(\varphi)&=&1 +\frac{1}{2}\beta_2\varphi^2+{\cal O}\left(\varphi^3\right),
\end{eqnarray}
where we have defined
$\beta_1\coloneqq A_{\varphi\varphi} (0)$ and
$\beta_2\coloneqq B_{\varphi\varphi} (0)$.
As at the background level the scalar field is trivial $\varphi=0$,
the Jordan and Einstein frames coincide, and $\tilde \rho=\rho$ and $\tilde p=p$.
For an incompressible star,
the Einstein field equations admit an exact solution of the
form~\eqref{eq:sss} given by~\cite{HTWW1965}
\begin{subequations}
\label{gr_exa}
\begin{align}
e^{\lambda(r)}&=\left(1-\frac{2GM r^2}{c^2R^3}\right)^{-1},
\\
e^{\nu(r)}
&=
\left[
\frac{3}{2}
 \left(1-\frac{2GM}{c^2R}\right)^{1/2}
-\frac{1}{2}
\left(1-\frac{2GM r^2}{c^2R^3}\right)^{1/2}
\right]^2,
\\
p(r)&=\rho c^2
\frac{\left(1-\frac{2GM r^2}{c^2R^3}\right)^{1/2}
      -\left(1-\frac{2GM}{c^2R}\right)^{1/2}}
      {3\left(1-\frac{2GM}{c^2R}\right)^{1/2}
        -\left(1-\frac{2GM r^2}{c^2R^3}\right)^{1/2}},
\end{align}
\end{subequations}
where $r=R$ is the surface of the star, at which $p(R)=0$.
Here, $M$ and ${\cal C}$ are the total mass and compactness of the star:
\begin{equation}
M=\frac{4\pi R^3}{3}{\rho},
\quad
{\cal C}=\frac{GM}{c^2R}.
\end{equation}

We then consider the perturbations to the background \eqref{gr_exa}
induced by the fluctuations of $\varphi$.
Since the corrections to the Einstein equations appear in
${\cal O}\left(\varphi^2,\varphi_\mu{}^2\right)$, at the leading order of $\varphi$
only the scalar field equation of motion becomes nontrivial. In the linearized
approximation, $\chi= 1+ {\cal O} (\varphi_\mu{}^2)$,
$\alpha= \beta_1\varphi+{\cal O}(\varphi^2)$
and
$\beta= \beta_2\varphi+{\cal O}(\varphi^2)$,
and the scalar field equation of motion \eqref{eq:scalar_eq}
for the massless and minimally coupled scalar field $P=2X$
reduces to
\begin{align}
\left(
g^{\rho\sigma}
-\frac{\kappa\Lambda}{2} T_{(\tm)}^{\rho\sigma}
\right)
\varphi_{\rho\sigma}
=-\frac{\kappa\beta_1}{2} T_{(\tm)}{}^\rho{}_\rho \varphi
+{\cal O}\left(\varphi^2,\varphi_\mu^2
\right).
\end{align}
Thus, as expected, in the Einstein frame
the corrections from disformal coupling appear
as the modification of the kinetic term
via the coupling to the energy-momentum tensor.

Taking the $s$-wave configuration for a stationary field,
$\dot{\varphi}=\ddot{\varphi}=0$, we get
\begin{align}
\varphi''
&+\frac{\frac{\nu'-\lambda'}{2}+\frac{2}{r}
-\frac{\kappa \Lambda}{2}
  \left[-\frac{\nu'}{2}\rho c^2+\left( -\frac{\lambda'}{2} +\frac{2}{r}\right)p(r)\right]}
{1-\frac{\kappa\Lambda}{2} {p(r)}}
\varphi'
\nonumber\\
&-\frac{\kappa \beta_1}{2} e^{\lambda(r)}
\frac{\rho c^2-3 p(r)}
      {1-\frac{\kappa\Lambda}{2} p(r)}\varphi
+{\cal O}\left(\varphi^2 ,\varphi'{}^2\right)
=0. 
\label{stationary_sca}
\end{align}

Inside the star, the scalar field equation of motion in the stationary
background~\eqref{stationary_sca} can be expanded as
\begin{align}
\label{pert_sca}
\varphi''
+\frac{2}{r}
\left[1+
{\cal O}\left({\cal C}\frac{r^2}{R^2}\right)
\right]
\varphi'
+u
\left[
1
+
{\cal O}\left({\cal C}\frac{r^2}{R^2}\right)
\right]
\varphi
=0,
\end{align}
where we have defined
\begin{eqnarray}
\label{u}
u\coloneqq
\frac{6\left(3\sqrt{1-2{\cal C} }-2\right){\cal C}}
       {\left(3\sqrt{1-2\cal C}-1\right)R^2+3 {\cal C}\left(\sqrt{1-2{\cal C}}-1\right)\Lambda}
      |\beta_1|.
      \nonumber \\
\end{eqnarray}
By neglecting the correction terms of order
${\cal O}\left({\cal C}\frac{r^2}{R^2}\right)$ in Eq.~\eqref{pert_sca},
the approximated solution inside the star
satisfying the regularity boundary condition at the center,
$\varphi(0)=\varphi_\tc$ and $\varphi'(0)=0$,
is given by
\begin{eqnarray}
\label{pert_in}
\varphi(r)\approx \varphi_\tc\frac{\sin (\sqrt{u}r)}{\sqrt{u}r}.
\end{eqnarray}
We note that at the surface of the star, $r=R$,
the corrections to this approximate solution \eqref{pert_in}
would be of ${\cal O}\left({\cal C}\right)$,
which is negligible for ${\cal C}\ll 1$
and gives at most a $10\%$ error even for ${\cal C}\simeq 0.1$.
Thus the solution \eqref{pert_in} provides a good approximation
to the precise interior solution of Eq.~\eqref{stationary_sca},
up to corrections of ${\cal O} (10\%)$ for typical NSs.

Outside the star, where $\tilde\rho=\tilde p=0$,
the scalar field equation of motion~\eqref{stationary_sca} reduces to
\begin{eqnarray}
\label{pert_sca2}
\varphi''
+
\left(
\frac{1}{r}
+\frac{1}{r-\frac{2GM}{c^2 }}
\right)
\varphi'
=
0.
\end{eqnarray}
The exterior solution of the scalar field is given by
\begin{eqnarray}
\label{pert_out}
\varphi(r)=
\varphi_0+
\frac{Q}{2M}
\ln
\left(
1-\frac{2GM}{c^2r}
\right),
\end{eqnarray}
which can be expanded as
\begin{eqnarray}
\varphi(r)=
  \varphi_0
-\frac{GQ}{c^2 r}
+{\cal O}\left(\frac{1}{r^2}\right),
\end{eqnarray}
where $Q$ denotes scalar charge. Matching at the surface $r=R$ gives
\begin{align}
\frac{GQ}{c^2R\varphi_0}
&=-\frac{2{\cal C}\left(1-2{\cal C}\right)\left(\sqrt{u}R-\tan(\sqrt{u}R) \right)}{\Xi},
\label{Qinf}
\\
\frac{\varphi_\tc}{\varphi_0}
&=-\frac{2{\cal C}\sqrt{u}R}{\cos\left(\sqrt{u}R\right)}
\frac{1}{\Xi},
\label{phi0}
\end{align}
where we introduced
\begin{align}
\Xi &= \left(1-2{\cal C}\right)\sqrt{u}R \ln \left(1-2{\cal C}\right) \nonumber \\
        &-\left[2{\cal C}+\left(1-2{\cal C}\right)\ln \left(1-2{\cal C}\right)\right]
        \tan (R\sqrt{u})
\end{align}

The scalar charge $Q$ and the central value of the scalar field $\varphi_\tc$
blow up when
\begin{eqnarray}
\frac{\tan\left(\sqrt{u}R\right)}{\sqrt{u}R}
=
\frac{(1-2{\cal C}) \ln (1-2{\cal C})}
      {2{\cal C} + \left(1-2{\cal C}\right)\ln (1-2{\cal C})}.
\end{eqnarray}
Thus, inside the star, the scalar field can be enhanced and
the scalarization takes place when
\begin{equation}
\label{enhance}
\sqrt{u}R\approx
\frac{\pi}{2}
\left(
1+\frac{4}{\pi^2}{\cal C}
\right).
\end{equation}
The condition~\eqref{enhance} can be rewritten as
\begin{eqnarray}
\label{scalarization}
|\beta^{\rm crit}_1|
&\approx&
\frac{\pi^2}{24{\cal C}}
\frac{3\sqrt{1-2\cal C}-1 +3{\cal C}\left(\sqrt{1-2{\cal C}}-1\right)\frac{\Lambda}{R^2}}
      {3\sqrt{1-2\cal C}-2}
\nonumber\\
&\times&
\left(1+\frac{4}{\pi^2}{\cal C}\right)^2,
\end{eqnarray}
where $\beta^{\rm crit}_1$ is the critical value of $\beta_1$ for which scalarization
can be triggered.

For small compactness ${\cal C}\ll 1$, we find at leading order
\begin{eqnarray}
|\beta^{\rm crit}_1|
\approx
\frac{\pi^2}{12{\cal C}}
\left(1 -\frac{3{\cal C}^2}{2R^2}\Lambda\right).
\label{eq:threfinal}
\end{eqnarray}
For a typical NS, the compactness parameter ${\cal C}\simeq 0.2$,
and if $\Lambda$ is negligibly small
$|\beta^{\rm crit}_1| = {\pi^2}/({12\,\cal C})\simeq 4.1$,
which agrees with the ordinary scalarization
threshold~\cite{Damour:1993hw,Harada:1997mr}.
On the other hand,
disformal coupling becomes important when $\Lambda\simeq (R/{\cal C})^2$,
which for $R\sim 10$ km and ${\cal C}\simeq 0.2$,
corresponds to $\Lambda\simeq 2500$ km$^2$.

In the other limit, for sufficiently large negative disformal coupling
parameters $|\Lambda|\gg (R/{\cal C})^2$, as
$uR^2\simeq 2 R^2/(|\Lambda|\, {\cal C}^2)\ll 1$,
from Eqs.~\eqref{Qinf} and \eqref{phi0} we have
\begin{equation}
\frac{GQ}{c^2R\varphi_0}\simeq -\frac{1-2{\cal C}}{3} uR^2\ll 1 \quad
{\rm and} \quad
\varphi_\tc\simeq \varphi_0,
\end{equation}
and the scalar field excitation is suppressed inside the star;
the stellar configuration is that of GR.

In the next section, we will show explicit examples of the numerical
integrations of the stellar structure and scalar field equations~[\eqref{set}
and~\eqref{eq:framedrag2}], and explore how the disformal coupling affects
the standard scalarization mechanism in the models proposed in
Refs.~\cite{Damour:1993hw,Damour:1996ke}.
We will confirm our main conclusions from the perturbative
calculations presented here.

\section{Numerical results}
\label{sec:results}

Having gained analytical insight into the effect of the disformal coupling on
spontaneous scalarization, we now will perform full numerical integrations of the
stellar structure equations.

For simplicity, we will focus on the simple case of a canonical scalar
field without a potential, $V(\varphi)=0$, and we will assume the
special form of the coupling
functions that enter Eq.~\eqref{eq:disformal}
\begin{equation}
A(\varphi)= e^{\frac{1}{2}\beta_1\varphi^2},\quad
B(\varphi)=e^{\frac{1}{2}\beta_2 \varphi^2},
\end{equation}
as a {\it minimal} model to include the disformal coupling in our problem.
In the absence of the disformal coupling function ($\Lambda = 0$), this model reduces to
that studied originally by Damour and
Esposito-Far\`ese~\cite{Damour:1993hw,Damour:1996ke}. Another input from the theory
is the cosmological value of the scalar field $\vp_0$, which for simplicity we take
to be zero  throughout this section. We also studied the case $\vp_0 = 10^{-3}$,
which does not alter our conclusions.

Under these assumptions our model is invariant under the
transformation $\vp \rightarrow -\vp$
(reflection symmetry). Therefore for each scalarized NS with scalar field configuration
$\vp$, there exists a reflection-symmetric counterpart with $\vp \rightarrow -\vp$.
For both families of solutions the bulk properties (such as masses, radii and moment
of inertia) are the same, while the scalar charges $Q$ have opposite sign, but the same
magnitudes. Moreover, $\vp = 0$ is a trivial solution of the stellar structure
equations. These solutions are equivalent to NSs in GR.

In this section we sample the ($\beta_1$, $\beta_2$, $\Lambda$)
parameter space of the theory, analyzing each parameter's influence on NS
models and on spontaneous scalarization. As mentioned in Sec.~\ref{sec:intro},
binary-pulsar observations have set a constraint of $\beta_1 \gtrsim -4.5$
in what corresponds to the purely conformal coupling ($\Lambda = 0$) limit of our
model. This lower bound on $\beta_1$ is not expected to apply for our more general
model and therefore, so far, the set of parameters ($\beta_1$, $\beta_2$, $\Lambda$)
are largely unconstrained.

\subsection{Equation of state}

To numerically integrate the stellar structure equations we must complement them
with a choice of EOS. Here we consider three realistic EOSs,
namely APR~\cite{Akmal:1998cf}, SLy4~\cite{Douchin:2001sv}
and FPS~\cite{Friedman:1981qw}, in decreasing
order of stiffness. The first two support NSs with masses larger than the
$M = 2.01 \pm 0.04 M_{\odot}$ lower bound from the pulsar PSR J0348+0432
in GR~\cite{Demorest:2010bx}. On the other hand, EOS FPS has a maximum mass of
$\sim 1.8 M_{\odot}$ in GR and is in principle ruled out by Ref.~\cite{Demorest:2010bx}.
Nevertheless, as we will see this EOS can support NSs with $M \gtrsim 2 M_{\odot}$,
albeit scalarized, for certain values of the theory's parameters.

With this set of EOSs we validated our numerical code by reproducing
the results of Refs.~\cite{Doneva:2014uma,Silva:2014fca} in the purely
conformal coupling limit. Our results including the presence of the
disformal coupling are presented next.

\subsection{Stellar models in the minimal scalar-tensor theory with disformal coupling}
\label{sec:numericalresults}

In Sec.~\ref{sec:incompress} we found that $\beta_1$ always needs to be
sufficiently negative for scalarization to be triggered. For this reason,
let us first analyze how $\Lambda$ and $\beta_2$ affect scalarized
nonrotating NSs assuming a fixed value of $\beta_1$.

\begin{figure}[t]
\begin{center}
\includegraphics[width=\columnwidth]{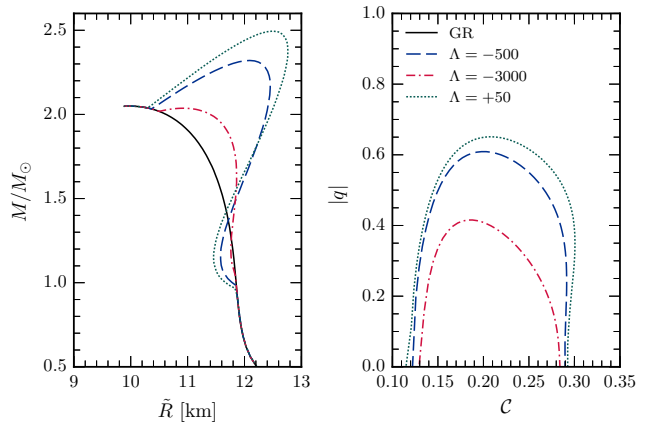}
\caption{We show the role of $\Lambda$ in spontaneous scalarization.
In both panels we consider stellar models using EOS SLy4
with $\beta_1 = -6.0$, $\beta_2 = 0$
and for $\Lambda = (-500,\, -3000,\, 50)$ km$^2$. For reference
the solid line corresponds to GR.
Left panel: The mass-radius relation.
Right panel: The dimensionless scalar charge $q \coloneqq -Q/M$~\cite{Damour:1993hw}
as a function of the compactness ${\cal C}$.
We see that $\Lambda > 0$ slightly increases scalarization
with respect to the purely conformal theory (cf. Fig.~\ref{fig:beta2}).
On the other hand, $\Lambda < 0$ can dramatically suppress
scalarization.
Note also that unlike $\beta_2$, $\Lambda$ can change the compactness
threshold above which scalarization can happen, as predicted by the analysis of
Sec.~\ref{sec:incompress}.
These results are qualitatively independent of the choice of EOS.
}
\label{fig:lambda}
\end{center}
\end{figure}

In Fig.~\ref{fig:lambda}, we consider what happens when we change the value of
$\Lambda$ while maintaining $\beta_1$ and $\beta_2$ fixed. We observe that for
sufficiently negative values of $\Lambda$ the effects of scalarization become
suppressed. This can be qualitatively understood from Eq.~\eqref{eq:threfinal}:
as $\Lambda/R^2 \rightarrow -\infty$ we need $|\beta^{\rm crit}_1| \rightarrow \infty$ for
scalarization to happen.
For fixed values of $\beta_1$ and ${\cal C}$,
there will be a sufficiently
negative value of $\Lambda$, for which $\beta^{\rm crit}_1 > \beta_1$ and
scalarization ceases to occur. Although in Fig.~\ref{fig:lambda} we show
$\Lambda = -3000$ km$^2$, we have confirmed this by constructing stellar models
for even smaller values of $\Lambda$. Also, in agreement with Sec.~\ref{sec:incompress},
we see that $\Lambda$ alters the threshold for scalarization. This is most clearly seen
in the right panel of Fig.~\ref{fig:lambda}, where for different values of $\Lambda$
scalarization starts (evidenced by a nonzero scalar charge $q$) when
different values of compactness ${\cal C}$ are
reached.\footnote{In the preceding section, because of the weak (scalar) field
approximation the Jordan and Einstein frame radii are approximately the same, i.e
${\tilde R} = R$. This is not the case in this section and hereafter the compactness
uses the Jordan frame radius, i.e. ${\cal C} = GM/(c^2 {\tilde R})$.}
In particular, for $\Lambda > 0$, because of the minus sign in the disformal term in
Eq.~\eqref{eq:threfinal}, NSs can scalarize for smaller values of ${\cal C}$,
while the opposite happens when $\Lambda < 0$.
We remark that for large positive $\Lambda$  the structure equations become singular
at the origin as discussed in Sec.~\ref{sec:case}. This prevents nonrelativistic stars, for which ${\cal C} \rightarrow 0$, from scalarizing.

\begin{figure}[htb]
\begin{center}
\includegraphics[width=\columnwidth]{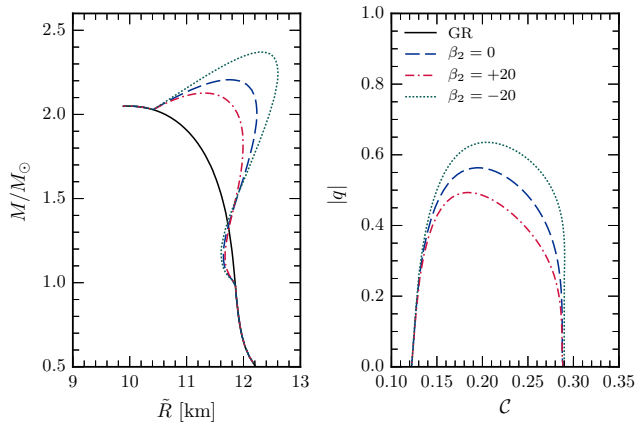}
\caption{We show the role of $\beta_2$ in spontaneous scalarization.
As in Fig.~\ref{fig:lambda}, in both panels we consider stellar models
using SLy4 EOS but with $\beta_1 = -6.0$ and
$\Lambda = -1000$ \kms for $\beta_2 = (-20,\, 0,\, 20)$. For reference the
solid line corresponds to GR.
Left panel: The mass-radius relation.
Right panel: The dimensionless scalar charge $q = -Q/M$
as a function of the compactness ${\cal C}$.
We see that $\beta_2$ affects highly scalarized stellar models making scalarization
stronger (in the sense of increasing the value of $q$) when $\beta_2 < 0$,
or weaker for $\beta_2 > 0$.
Observe that $\beta_2$ has a negligible effect on weakly scalarized models
($|q|\lesssim 0.35$). This is in agreement with its absence from
the perturbative analysis of Sec.~\ref{sec:incompress}.
Note that the range of ${\cal C}$ for which scalarization occurs is
the same, irrespective of the choice of $\beta_2$.
Again, these results are qualitatively independent of the choice of EOS.
}
\label{fig:beta2}
\end{center}
\end{figure}

In Fig.~\ref{fig:beta2}, we consider what happens when we change the value of
$\beta_2$ while maintaining $\beta_1$ and $\Lambda$ fixed. We see that in agreement
with Eq.~\eqref{eq:threfinal}, the parameter $\beta_2$ does not affect the threshold
for scalarization. Moreover, we observe that $\beta_2 < 0$ ($\beta_2 > 0$)
makes scalarization more (less) evident with respect to $\beta_2 = 0$. In fact, in
Eqs.~\eqref{eq:c2} and~\eqref{eq:c1}, we see that $\beta_1$ and $\beta_2$ contribute
to the scalar field equation through the factors $\Lambda A^4 B^2$ and
$\beta - \alpha$, which have competing effects in sourcing the scalar field for
$\beta_1 < 0$ and $\beta_2 \neq 0$. Our numerical integrations indicate
that the former is dominant and that $\beta_2 \neq 0$ affects only very compact
NSs (${\cal C}\gtrsim 0.15$ in the example of Fig.~\ref{fig:beta2}).

\begin{figure}[h]
\begin{center}
\includegraphics[width=\columnwidth]{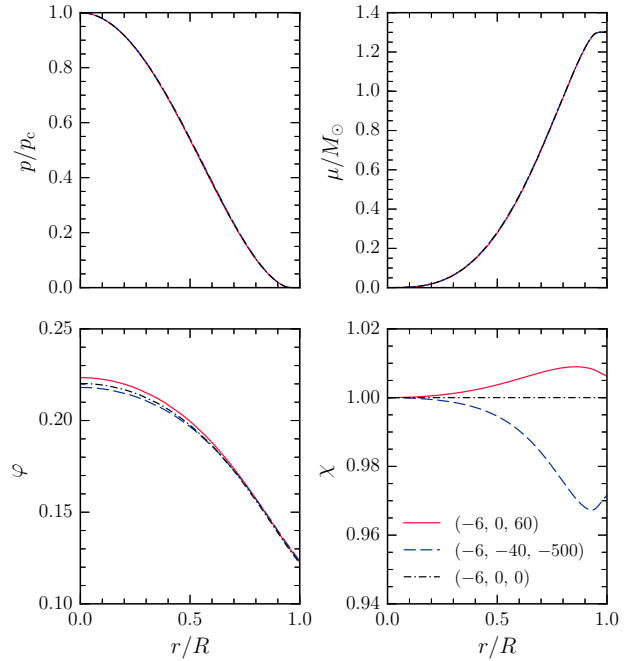}
\caption{
We show the normalized pressure profile $p/p_{\rm c}$ (top left),
dimensionless mass function $\mu/M_{\odot}$
(top right), scalar field $\varphi$ (bottom left) and
the disformal factor $\chi$ (bottom right) in the stellar interior.
The radial coordinate was normalized by the Einstein frame radius $R$.
The radial profiles above correspond to three stellar configurations using
SLy4 EOS, with fixed baryonic mass $M_\tb/M_{\odot} = 1.5$ and theory
parameters $(\beta_1, \beta_2, \Lambda)$ = $(-6, 0, 60)$, $(-6, -40, -500)$
and $(-6, 0, 0)$, the latter corresponding to a stellar model in the
Damour-Esposito-Far\`ese theory~\cite{Damour:1993hw,Damour:1995kt}.
While the fluid variables are not dramatically affected, models with
$\Lambda > 0$ ($\Lambda < 0$) become more (less) scalarized due to the
disformal coupling. The bulk properties of these models are
summarized in Table~\ref{tab:models}.
}
\label{fig:prof}
\end{center}
\end{figure}

It is also of interest to see how scalarization affects the interior of NSs.
In Fig. \ref{fig:prof}, we show the normalized pressure profile $p/p_{\rm c}$
(top left), the dimensionless mass function $\mu/M_{\odot}$ (top right),
the scalar field $\varphi$ (bottom left) and
the disformal factor $\chi$ (bottom right) in the stellar interior.
The radial coordinate was normalized by the Einstein frame radius $R$.
The quantities correspond to three stellar configurations using SLy4 EOS
with fixed baryonic mass $M_\tb / M_{\odot} = 1.5$, which in GR yields a
canonical NS with mass $M \approx 1.4\, M_{\odot}$, for the sample values of
$(\beta_1, \beta_2, \Lambda)$ indicated in Table~\ref{tab:models}.
In agreement with our previous discussion we see that NSs with $\Lambda>0$ ($\Lambda<0$)
support a larger (smaller) value of $\vp_\tc$, which translates to a
larger (smaller) value of $q$. It is particularly important to observe that
$\chi$ is non-negative for all NS models, guaranteeing the Lorentzian signature
of spacetime [cf. Eq.~\eqref{eq:chi}].

\begin{table}
\begin{tabular}{c c c c c c }
\hline
\hline
$(\beta_1,\, \beta_2,\, \Lambda)$  &  $\tilde{R}$ [km]  &  $M$ [$M_{\odot}$]  & $I$ [$10^{45}$g cm$^2$]
&$\varphi_{\rm c}$ & $q$\Tstrut\Bstrut \\
\hline
GR & 11.72 & 1.363 & 1.319 & -- & -- \\
$(-6, 0, 0)$  &  11.60  &  1.354  & 1.431 & 0.220  &  0.613\\
$(-6, -40, -500)$  &  11.64  &  1.354  &  1.438 & 0.218 &  0.622 \\
$(-6, 0, 60)$  &  11.59  &  1.354  &  1.430 & 0.223 &  0.615 \\
\hline
\hline
\end{tabular}
\caption{The properties of NSs in GR and scalar-tensor theory
using EOS SLy4 and fixed baryonic mass $M_{\rm b}/M_{\odot} = 1.5$.
The radial profiles of some of the physical variables involved in
the integration of the stellar model are shown in Fig.~\ref{fig:prof}.
}
\label{tab:models}
\end{table}

\begin{figure*}[htb]
\begin{center}
\includegraphics[width=1.9\columnwidth]{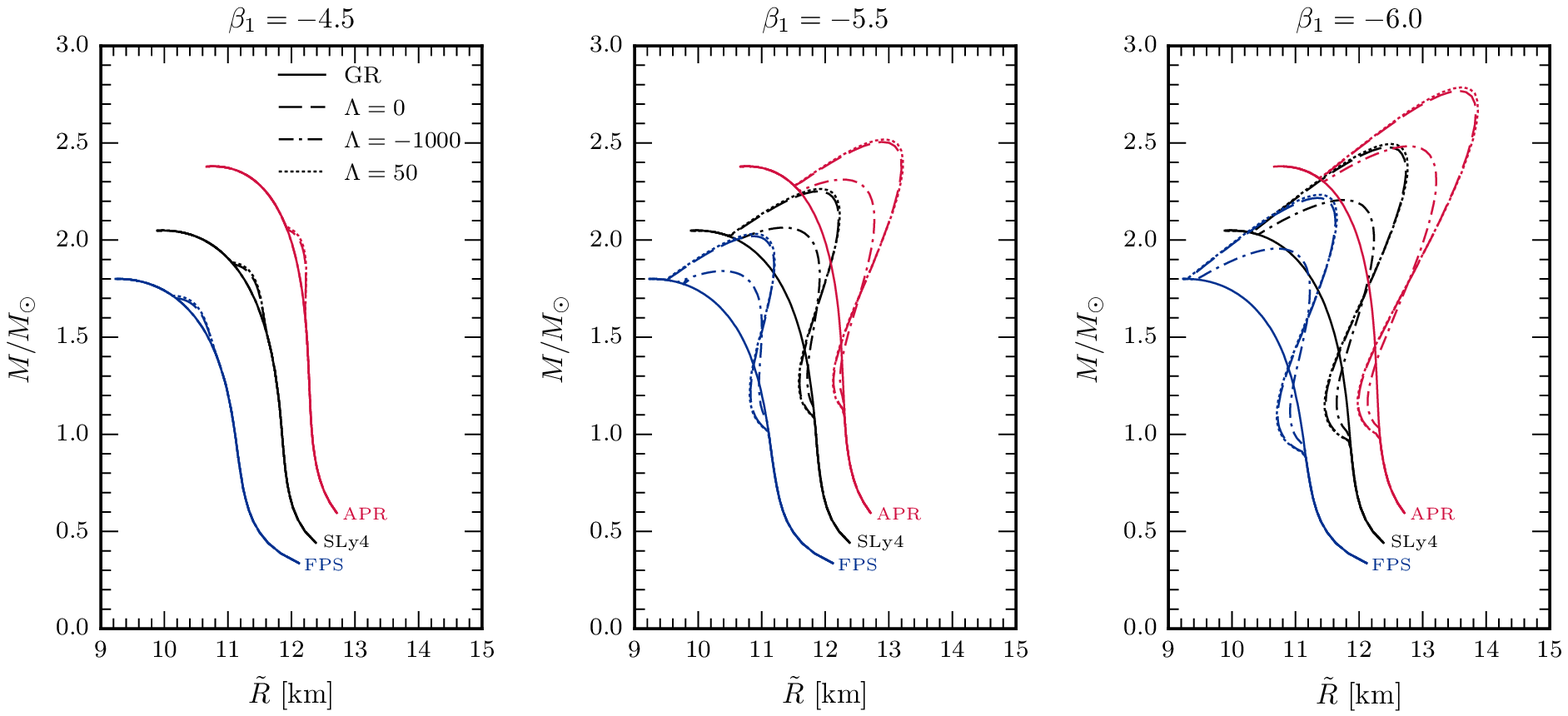}
\includegraphics[width=1.9\columnwidth]{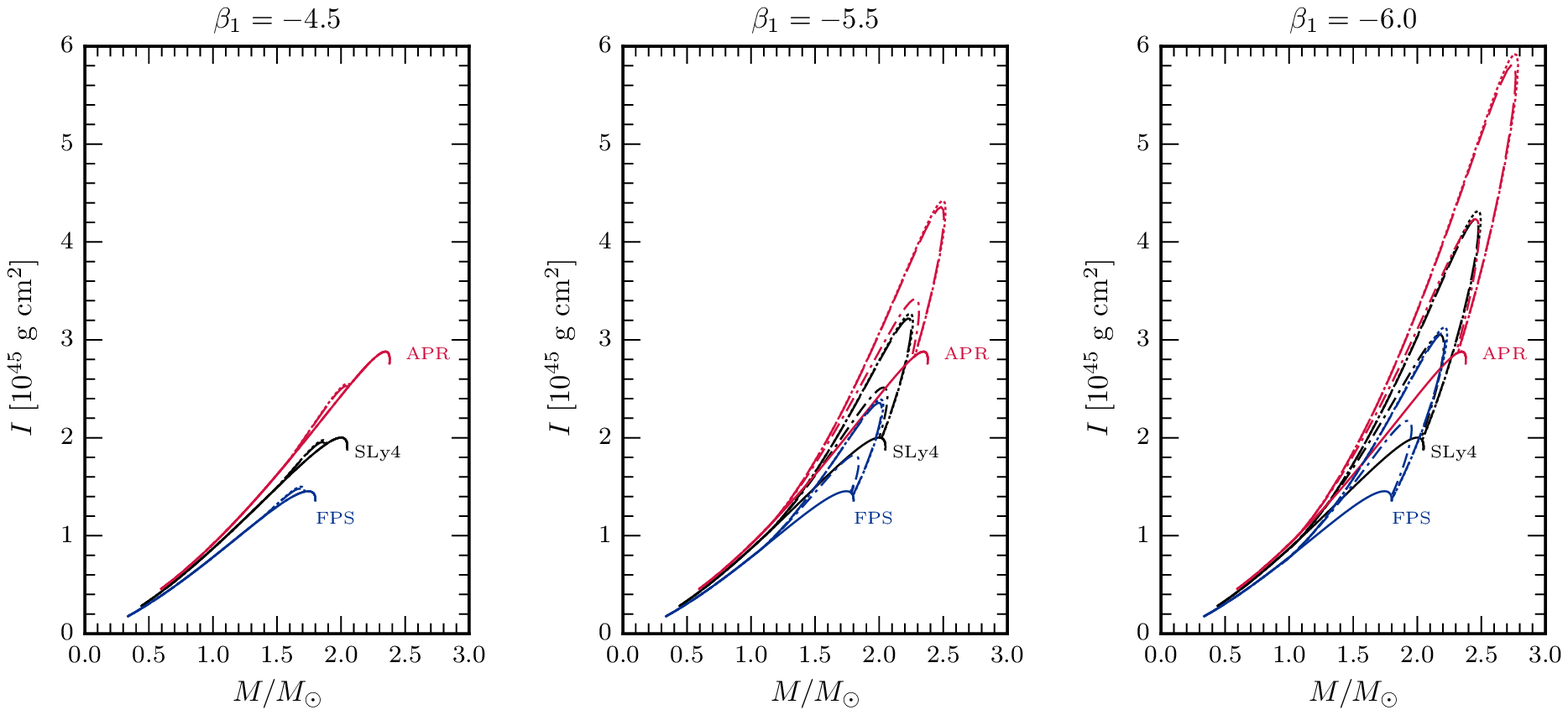}
\caption{We show NS models in scalar-tensor theories with disformal coupling
for three choices of realistic EOSs, namely APR, SLy4 and FPS, in decreasing
order of stiffness. We illustrate the effect of varying the values of $\beta_1$ and
$\Lambda$, while keeping $\beta_2$ fixed ($\beta_2 = 0$) for simplicity. The
curves corresponding to $\Lambda = 0$, represent stellar models in purely
conformal theory~\cite{Damour:1993hw,Damour:1996ke}.
Top panels: Mass-radius relations. Bottom panels: Moment of inertia versus mass.
As seen in Fig.~\ref{fig:lambda} already, while $\Lambda < 0$ weakens scalarization,
$\Lambda > 0$ strengthens the effect. For $\beta_2 = 0$, this latter effect
is very mild, being more evident by $\beta_2 < 0$ (cf. Fig.~\ref{fig:massradiusb2}).
}
\label{fig:massradius}
\end{center}
\end{figure*}

In Fig.~\ref{fig:massradius} we show the mass-radius curves (top panels) and
moment of inertia-mass (lower panels) for increasing values of $\beta_1$
(from left to right), for three realistic EOSs, keeping $\beta_2 = 0 $, but using
different values of $\Lambda$. As we anticipated in Fig.~\ref{fig:lambda},
negative values of $\Lambda$ reduce the effects of
scalarization, while positive values increase them.
The case $\Lambda = 0$ corresponds to the purely conformal theory
of Ref.~\cite{Damour:1993hw}. We observe that scalarized NS models
branch from the GR family at different points for different values of
$\Lambda$ (when $\beta_1$ is fixed).
In agreement with our previous discussion, sufficiently
negative values of $\Lambda$ can completely suppress scalarization.
Indeed for
$\beta = -4.5$ the solutions with $\Lambda = -1000$ km$^2$ are identical to GR, while
scalarized solutions exist when $\Lambda = 0$.

Additionally, we observe degeneracy between families of solutions in
theories with different parameters. For instance, the maximum mass for a NS assuming
EOS APR is approximately the same, $M/M_{\odot} \approx 2.38$,
for both $\beta_1 = -5.5$, $\Lambda = 0$ and $\beta_1 = -6.0$, $\Lambda = -1000$ km$^2$.
We also point out the degeneracy between the choice of EOS
and of the parameters of the theory.
For instance, the maximum mass predicted by EOS FPS
in the theory with $\beta_1 = -5.5$ and $\Lambda = 50$ km$^2$ is approximately
the same as that predicted by GR, but for EOS SLy4, i.e $M/M_{\odot} \approx 2.05$.
We emphasize that these two types of degeneracies are not exclusive to the theory we are considering, but are generic to {\it any} modification to GR~\cite{Glampedakis:2015sua}.

\begin{figure*}[htb]
\begin{center}
\includegraphics[width=1.9\columnwidth]{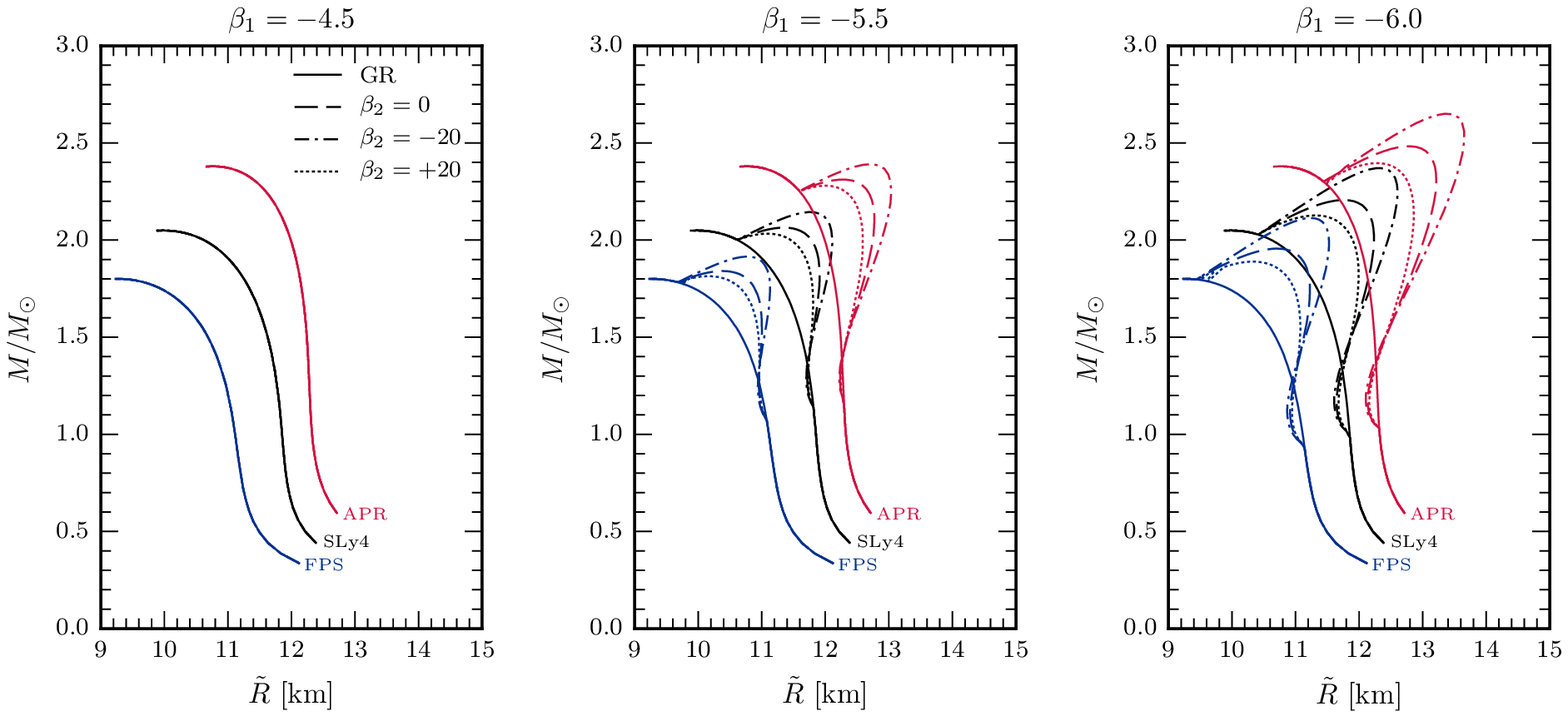}
\includegraphics[width=1.9\columnwidth]{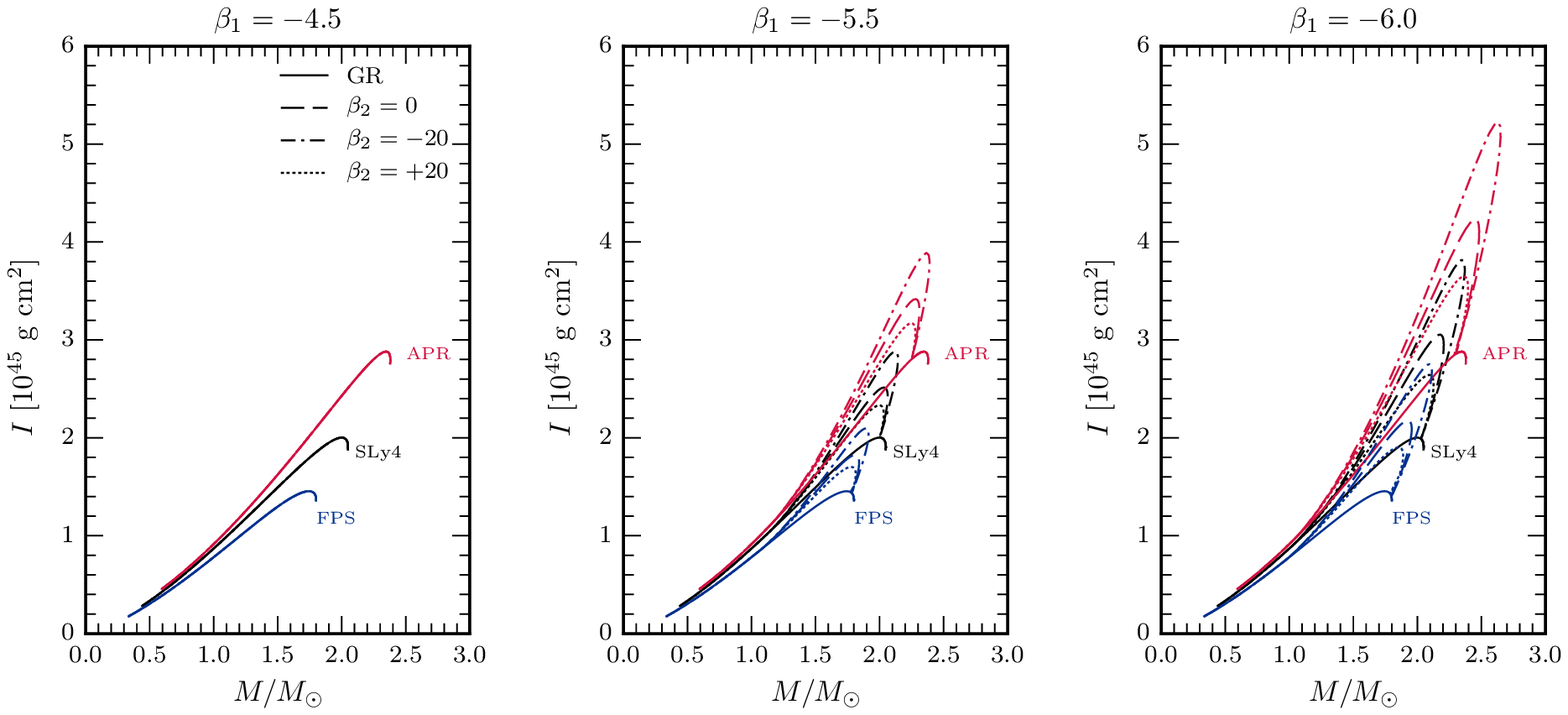}
\caption{
In comparison to Fig.~\ref{fig:massradius}, here we show the influence of $\beta_2$
in spontaneous scalarization while keeping $\Lambda = -1000$ km$^2$. As we
have seen in Fig.~\ref{fig:lambda} (and by the analytic treatment of
Sec.~\ref{sec:incompress}), negative values of $\Lambda$ suppress scalarization.
This effect is such that for $\beta_1 = -4.5$, scalarization is
suppressed altogether (top left panel). For smaller values of $\beta_1$,
this value of $\Lambda$ weakens scalarization and we clearly see that
$\beta_2$ affects the most scalarized stellar models in the conformal
coupling theory. Note that the range covered by the axis here and in
Fig.~\ref{fig:massradius} is the same, making it clear that scalarization is
less strong for the values of $\beta_2$ adopted.
}
\label{fig:massradiusb2}
\end{center}
\end{figure*}

In Fig.~\ref{fig:massradiusb2}, we exhibit the mass-radius (top panels) and
moment of inertia-mass (lower panels) for increasing values of $\beta_1$
(from left to right), but now keeping $\Lambda = -1000$ km$^{2}$ and changing
the value of $\beta_2$. Once more, sufficiently negative values of $\Lambda$ can
completely suppress scalarization. This is clearly seen in the panels for
$\beta_1 = -4.5$, where $\Lambda = -1000$ km$^2$, suppresses scalarization for all
values of $\beta_2$ considered. We observe that independently of the choice of EOS,
$\beta_2 > 0$ ($\beta_2 < 0$) yields smaller (larger) deviations from GR.

\subsection{Stability of the solutions}
\label{sec:stability}

\begin{figure}[htb]
\begin{center}
\includegraphics[width=\columnwidth]{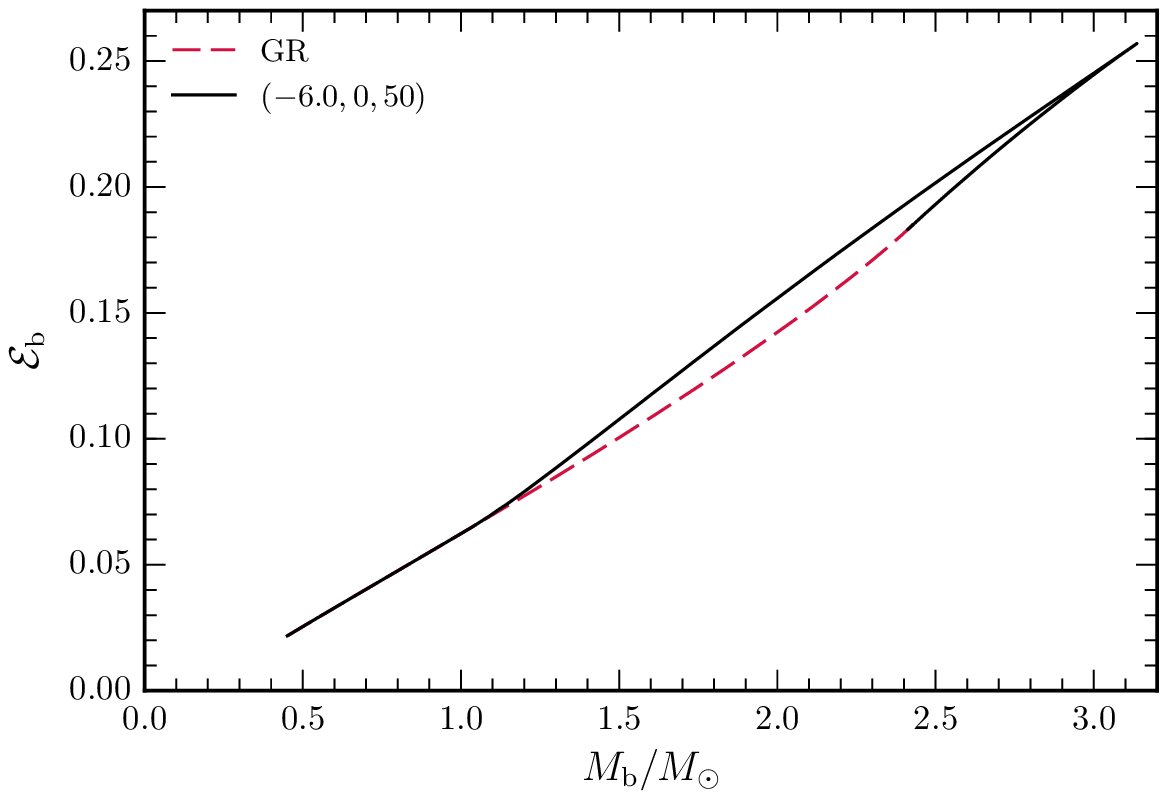}
\caption{We show the fractional binding energy ${\cal E}_\tb$ as a function
of the baryonic mass for stellar models using EOS SLy4 and for theory
with $(\beta_1, \beta_2, \Lambda) = (-6,0,50)$. Solutions in this theory branch
around $M_{\rm b} / M_{\odot} \approx 1.1$ with scalarized solutions (solid line)
being energetically favorable over the general-relativistic ones (dashed line). The
turning point at the solid curve corresponds to the maximum in the $M$-$R$ relation,
cf. Fig.~\ref{fig:lambda}.
}
\label{fig:energ}
\end{center}
\end{figure}

Let us briefly comment on the stability of the scalarized solutions obtained
in this section. In general, for a given set of parameters
$(\beta_1, \beta_2, \Lambda)$ and fixed values of
$M_{\rm b}$ and $\vp_{0}$, we have more than one stellar configuration
with different values of the mass $M$. Following the arguments
of Refs.~\cite{Damour:1993hw,Harada:1997mr,Horbatsch:2010hj}, we take the solution
of smallest mass $M$, i.e., larger fractional
binding energy ${\cal E}_\tb$ defined in Eq.~\eqref{eq:binding},
to be the one which is energetically favorable to be realized in nature.
In Fig.~\ref{fig:energ}, we show ${\cal E}_{\rm b}$ as a function of $M_{\rm b}$
for the two families of solutions in a theory with $(\beta_1, \beta_2, \Lambda)
= (-6,0,50)$ and $\vp_0 = 0$. The dashed line corresponds to solutions which
are indistinguishable from the ones obtained in GR, while the solid line
(which branches off from the former around $M_{\rm b}/M_{\odot} \approx 1.1$)
corresponds to scalarized solutions. We see that scalarized stellar
configurations in our model are energetically favorable, as happens in the case
of purely conformal coupling theory~\cite{Damour:1993hw,Harada:1997mr,Horbatsch:2010hj}.


\section{An application: EOS-independent $I$-${\cal C}$ relations}
\label{sec:icrelations}

As we have seen in the previous sections the presence of the disformal
coupling modifies the structure of NSs making scalar-tensor theories
generically predict different bulk properties with respect to GR. However,
as we discussed based on Figs.~\ref{fig:massradius} and~\ref{fig:massradiusb2},
modifications caused by scalarization are usually degenerate with the choice of EOS,
severely limiting our ability to constrain the parameters of the theory using
current NS observations (see e.g. Ref.~\cite{He:2014yqa}).
Moreover, different theory parameters can yield similar stellar models for a fixed EOS.

An interesting possibility to circumvent these problems is to search
for EOS-independent (or at least weakly EOS-dependent) properties of NSs.
Accumulating evidence favoring the existence of such EOS independence between
certain properties of NSs, culminated with the discovery of the
$I$-Love-$Q$ relations~\cite{Yagi:2013bca,Yagi:2013awa} connecting
the moment of inertia, the tidal Love number and the rotational quadrupole moment
(all made dimensionless by certain multiplicative factors) of NSs in GR.

If such relations hold in modified theories of gravitation they can potentially
be combined with future NS measurements to constrain competing theories of gravity.
This attractive idea was explored in the context of
dynamical Chern-Simons theory~\cite{Yagi:2013awa},
Eddington-inspired Born-Infeld gravity~\cite{Sham:2013cya},
Einstein-dilaton-Gauss-Bonnet (EdGB) gravity~\cite{Kleihaus:2014lba,Kleihaus:2016dui},
$f(R)$ theories~\cite{Doneva:2015hsa} and the Damour-Esposito-Far\`ese
model of scalar-tensor gravity~\cite{Doneva:2013qva,Pani:2014jra}.

Within the present framework we cannot compute the $I$-Love-$Q$ relations,
since while on one hand we can compute $I$, the tidal Love
number requires an analysis of tidal interactions, and
the rotational quadrupole moment
$Q$ requires pushing the Hartle-Thorne perturbative expansion up
to order ${\cal O}(\Omega^2)$. Nevertheless,
we can investigate whether the recently proposed $I$-${\cal C}$
relations~\cite{Breu:2016ufb} between the moment of inertia $I$ and the
compactness ${\cal C}$ remain valid in our theory.
For a recent study in the Damour-Esposito-Far\`ese and $R^2$ theories,
see Ref.~\cite{Staykov:2016mbt}. This relation was also studied
for EdGB and the subclass of Horndeski gravity with
nonminimal coupling between the scalar field and the Einstein tensor
in Ref.~\cite{Maselli:2016gxk}.

The relation proposed in Ref.~\cite{Breu:2016ufb} for the moment of inertia
$\bar{I} \coloneqq I/M^3$ and the compactness ${\cal C}$
is
\begin{equation}
\bar{I} = a_{1}\, {\cal C}^{-1} + a_{2}\, {\cal C}^{-2}
+ a_{3}\, {\cal C}^{-3} + a_{4}\, {\cal C}^{-4}\,,
\label{eq:icbreurez}
\end{equation}
where the coefficients $a_{i}$ ($i = 1, \ldots, 4$) are given by
$a_1 = 8.134 \times 10^{-1}$, $a_2 = 2.101 \times 10^{-1}$,
$a_3 = 3.175 \times 10^{-3}$ and $a_4 = -2.717 \times 10^{-4}$. This
result is valid for slowly rotating NSs in GR, although it
can easily be adapted for rapidly rotating NSs~\cite{Breu:2016ufb}.
The coefficients in Eq.~\eqref{eq:icbreurez} are obtained by
fitting the equation to a large sample of EOSs. For earlier work
considering a different normalization for $\bar{I}$, namely
$I/(M R^2)$, see e.g. Refs.~\cite{1994ApJ...424..846R,Lattimer:2000nx,Bejger:2002ty,Lattimer:2004nj,Urbanec:2013fs}.

\begin{figure}
\begin{center}
\includegraphics[width=\columnwidth]{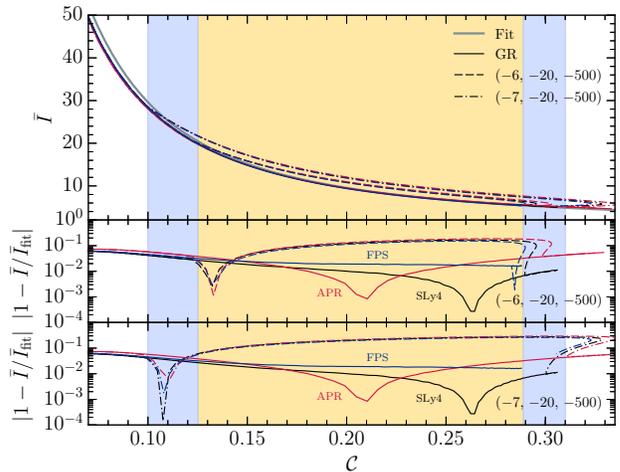}
\caption{We consider the $I$-${\cal C}$ relation in scalar-tensor gravity.
Top panel: The fit \eqref{eq:icbreurez} obtained in the context of GR (thick solid line) is confronted against stellar models obtained in GR (solid line); and scalar-tensor
theories with parameters $(\beta_2, \Lambda) = (-20, -500)$, but with
$\beta_1 = -6$ (dashed lines) and $\beta_2 = -7$ (dash-dotted lines), using EOSs
APR, SLy4 and FPS.
Middle panel: Relative error between the fit for GR against scalar-tensor theory
with $\beta_1 = -6$.
Bottom panel: Similarly, but for $\beta_1 = -7$.
In all panels the shaded regions correspond to approximately
the domain of compactness for which spontaneous scalarization occurs in each
theory. While for GR, the errors are typically below $6 \%$, scalarized models can
deviate from GR by $20\%$ (for $\beta_1 = -6$) and up to
$40\%$ (for $\beta_1 = -7$).
}
\label{fig:icrel}
\end{center}
\end{figure}

We confront this fit against stellar models in two scalar-tensor theories
with the parameters $(\beta_1, \beta_2, \Lambda)$ having the values
$(-6, -20, -500)$ and $(-7, -20, -500)$ that support highly scalarized solutions.
As seen in Fig.~\ref{fig:icrel}, the deviations from GR
can be quite large, up to $40\%$ for the theory with $\beta_1 = -7$ in the
range of compactness for which spontaneous scalarization happens
(cf. Fig.~\ref{fig:icrel}, bottom panel). Nevertheless, the EOS
independence between $\bar{I}$ and ${\cal C}$ remains even
when scalarization occurs (cf. Fig.~\ref{fig:icrel}, top panel).

Since our model is largely unconstrained observationally, measurements
of the moment of inertia and compactness of NSs could in principle be used to
constrain it or, more optimistically, indicate the occurrence of spontaneous
scalarization in NSs. This is in contrast with the standard
Damour-Esposito-Far\`ese model, for which the theory's parameters are so
tightly constrained by binary pulsar observations~\cite{Antoniadis:2012vy},
that spontaneous scalarization (if it exists) is bound to have a negligible
influence on the $I$-${\cal C}$ relation~\cite{Staykov:2016mbt}.
We stress however that in general it will be difficult to constrain the parameter
space $(\beta_1, \beta_2, \Lambda)$ only through the $I$-${\cal C}$ relation.
The reason is in the degeneracy of stellar models for different values
of the parameters; see the discussion in Sec.~\ref{sec:numericalresults}.

\section{Conclusions and outlook}
\label{sec:conclusions}

In this paper we have presented a general formulation
to analyze the structure of relativistic stars
in scalar-tensor theories with disformal coupling,
including the leading-order corrections due to slow rotation.
The disformal coupling is negligibly small in comparison with conformal
coupling in the weak-gravity or slow-motion regimes,
where the scalar field is slowly evolving and
typical pressures are much smaller than the energy density scales,
but it may be comparable to the ordinary conformal coupling in
the strong-gravity regime found inside relativistic stars.
Our calculation covers a variety of scalar-tensor models,
especially, conformal and disformal couplings to matter,
nonstandard scalar kinetic terms and generic scalar potential terms.

After obtaining the stellar structure equations,
we have particularly focused on the case of a
canonical scalar field with a generic scalar potential. We showed that
in the absence of both conformal coupling and a scalar potential,
the disformal coupling does not modify the stellar structure with respect to GR.
On the other hand, this result shows us that inside relativistic stars
the effects of disformal coupling always appear {\it only} when there is
conformal coupling to matter and/or a nontrivial potential term. The strength
of disformal coupling crucially depends on the coupling strength $\Lambda$
in Eq.~\eqref{eq:disformal} with dimensions of (length)$^2$. For a canonical scalar
field, $\Lambda$ has to be of ${\cal O}(10^3)$ km$^{2}$ to significantly
influence the structure of NSs.

In our numerical analyses, we have investigated the effects of the
disformal coupling on the spontaneous scalarization of NSs in the scalar-tensor
theory with purely conformal coupling. We found that the effects of disformal
coupling depend on the sign of $\Lambda$. We showed that for negative values of
$\Lambda$ the mass and moment of inertia of NSs decrease, approaching the values
in GR for sufficiently large negative values of $\Lambda$.
We speculate that this is the consequence of a mechanism similar to the
disformal screening proposed in Ref.~\cite{Koivisto:2012za} where in a high
density or a large disformal coupling limit the response of the scalar field
becomes insensitive to the local matter density, exemplified here by studying
relativistic stars.
On the other hand, for positive values of $\Lambda$, we showed that the mass and
moment of inertia increase but for too large positive values of $\Lambda$ the
stellar structure equation becomes singular and a regular NS solution
cannot be found. This allowed us to derive a mild upper bound of
$\Lambda \lesssim 100$ km$^2$, that does not depend on the choice of the EOS.

We have also tested the applicability of a recently proposed
EOS-independent relation between the dimensionless moment of inertia $I/M^3$
and the compactness ${\cal C}$ for NSs in GR. We found that for a certain domain
of the theory's parameter space, the deviations from GR can be as large as
$\sim 40\%$, suggesting that future measurements of NS moment of inertia might be
used to test scalar-tensor theories with disformal coupling. Because of the large
dimensionality of the parameter space, modifications with respect to GR are
generically degenerate between different choices of $\beta_1$, $\beta_2$ and
$\Lambda$. Thereby, even though deviations from GR can be larger, it seems
unlikely that constraints can be put on the theory's parameters
using exclusively the $I$-${\cal C}$ relation. In this regard, it would be
worth extending our work and studying how the $I$-Love-$Q$ relations are affected
by the disformal coupling, generalizing the works of
Refs.~\cite{Pani:2014jra,Doneva:2014faa,Doneva:2013qva} for scalar-tensor theories
with disformal coupling.

Still in this direction, one could investigate whether the
``three-hair'' relations -- EOS-independent relations connecting higher-order
multipole moments of rotating NSs in terms of the first three multipole moments
in GR~\cite{Pappas:2013naa,Yagi:2014bxa,Majumder:2015kfa} - remain valid in
scalar-tensor theory, including those with disformal coupling. This could be
accomplished by combining the formalism developed in~\cite{Pappas:2014gca}
with numerical solutions for rotating NSs such as those obtained
in Ref.~\cite{Doneva:2013qva}.

Although the main subject of this paper was to investigate the hydrostatic
equilibrium configurations in scalar-tensor theories with disformal coupling,
let us briefly comment on the gravitational (core) collapse resulting in the
formation of a NS (see e.g. Ref.~\cite{Gerosa:2016fri}). A fully numerical analysis
of dynamical collapse in this theory is beyond the scope of our paper, but an
important issue in this dynamical process may be the possible appearance of
ghost instabilities for negative values of $\Lambda$~\cite{Kaloper:2003yf,Koivisto:2012za,Berezhiani:2013dw,Bettoni:2015wta}.
During collapse, matter density at a given position increases, and if at some
instant it reaches the threshold value where the effective kinetic term in the scalar field equation vanishes, the time evolution afterwards cannot be determined. For a
canonical scalar field $P=2X$, in a linearized approximation where $\chi\simeq 1$
and $B(\varphi)\simeq 1$, the effective kinetic term of the equation of
motion~\eqref{eq:scalar_eq} is roughly given by
\begin{align}
-\left(1-\frac{\kappa|\Lambda|}{2}\tilde\rho c^2\right)\ddot{\varphi},
\end{align}
where a dot represents a time derivative. The sign of the kinetic term may change
in the region of a critical density higher than
${\tilde \rho}_{\rm crit}= {2}/\left({\kappa c^2|\Lambda|}\right)$. The choice of
$\Lambda=-100$ km$^2$ gives $\tilde \rho_{\rm crit} \simeq  10^{15}$ g/cm$^3$, which
is a typical central density of NSs. Thus for $|\Lambda| \lesssim 100$ km$^2$ a NS
is not expected to suffer an instability while for other values it might occur in the
interior of the star.
Of course, for a more precise estimation, nonlinear interactions between the
dynamical scalar field, spacetime and matter must be taken into consideration.
A detailed study of time-dependent processes in our theory is definitely
important, but is left for future work.

Another interesting prospect for future work would be to study compact binaries
within our model. The most stringent test of scalar-tensor gravity comes from the
measurement of the orbital decay of binaries with asymmetric
masses, which constrains the emission of dipolar scalar radiation by the
system~\cite{Freire:2012mg}. We expect that the disformal coupling parameters
$\beta_2$ and $\Lambda$ should play a role in the orbital evolution of a
binary system by influencing the emission of scalar radiation from the system.
In fact, both parameters are expected to modify the so-called
sensitivities~\cite{Will:1989sk,Zaglauer:1992bp} that enter at the lowest
PN orders sourcing the emission of dipolar scalar radiation.
An investigation of compact binaries within our model
could, combined with current observational data, yield
tight constraints on disformal coupling.
Moreover one could study NS solutions for other classes
of scalar-tensor theories not considered here. This task is facilitated by the
generality of our calculations presented in Sec.~\ref{sec:structure}.
Work in this direction is currently underway and we hope to report it soon.

\section*{Acknowledgements}
We would like to thank Andrea Maselli, Emanuele Berti and George Pappas for
suggestions on our draft. We thank Jeremy Sakstein and Kent Yagi
for several interesting comments and Caio F. B. Macedo for discussions during
the development of this work.
%
%
We also thank the anonymous referee for important suggestions and insightful
comments on this work.
H.O.S. thanks the hospitality of the Instituto Superior T\'ecnico (Portugal)
during the final stages of preparation of this work.
This work was supported by FCT-Portugal through Grant No. SFRH/BPD/88299/2012 (M.M.)
and a NSF CAREER Grant No. PHY-1055103 (H.O.S.).

\appendix

\section{Disformal invariance}
\label{sec:inv}

In this appendix, we study how the physical quantities associated with the
stellar properties transform under the disformal
transformation~\eqref{eq:disformal}. We write the metrics with slow rotation of
spacetimes in the Einstein and Jordan frames as
\begin{align}
\label{ae}
ds^2 &= -e^{\nu(r)} c^2 dt^2
+ e^{\lambda(r)} dr^2 + r^2 \left( d\theta^2
+ \sin^2\theta d\phi^2 \right)
 \nonumber \\
&+2\left(\omega-\Omega\right)
r^2 \sin^2\theta dt d\phi,
\end{align}
and
\begin{align}
\label{aj}
d{\tilde s}^2 &= -e^{\bar \nu(\bar r)}
c^2 dt^2 + e^{\bar \lambda(\bar r)} d{\bar r}^2
+{\bar  r}^2 \left( d\theta^2
+ \sin^2\theta d\phi^2 \right)
 \nonumber \\
&+2\left(\bar \omega-\bar \Omega\right)
{\bar r}^2 \sin^2\theta dt d\phi.
\end{align}
We can relate Eqs. \eqref{ae} and \eqref{aj} using the disformal
relation~\eqref{eq:disformal} as
\begin{eqnarray}
\label{f1}
e^{\bar \nu}&=& A^2(\varphi) e^\nu,
\\
\label{f2}
e^{\frac{\bar \lambda}{2}} d{\bar r}&=&A(\varphi)\sqrt{\chi} e^{\frac{\lambda}{2}} dr,
\\
\label{f3}
{\bar r}&=& r A(\varphi),
\\
\label{f4}
{\bar \omega}-{\bar \Omega}
&=&\omega-\Omega,
\end{eqnarray}
where we recall that due the symmetries of the problem $\vp = \vp(r)$.
From Eqs.~\eqref{f2} and \eqref{f3} we get
\begin{eqnarray}
\label{f5}
e^{\bar \lambda}
=\frac{\chi}{(1+r\alpha\varphi')^2}e^\lambda.
\end{eqnarray}
Introducing $\mu$ and ${\bar\mu}$ in the Einstein and Jordan frames by
\begin{eqnarray}
e^{-\lambda}= {1-\frac{2\mu}{r}},
\quad
e^{-{\bar \lambda}}={1-\frac{2\bar \mu}{\bar r}},
\end{eqnarray}
and using Eqs.~\eqref{f3} and \eqref{f5} we find
\begin{eqnarray}
\label{bmr}
\bar \mu
=-\frac{r A(\varphi)}{2}
\left[
\left(1-\frac{2\mu}{r}\right)
\frac{(1+r\alpha(\varphi)\varphi')^2}{\chi}-1
\right].
\end{eqnarray}
As it is reasonable to set $\varphi_0=0$ and $\varphi_0'=0$ at asymptotic
infinity, in the class of models considered in the text [Eq. \eqref{def:ab}],
$A(\varphi_0)=1$, $\alpha(\varphi_0)=0$ and $\chi(\varphi_0,\varphi_0')=1$,
we find that the ADM mass obtained from the leading-order values of $\mu$ and
$\bar\mu$ at asymptotic infinity is disformally invariant
\begin{eqnarray}
\label{adm_equiv}
\bar M=M.
\end{eqnarray}

The energy-momentum tensors of the matter fields in the Einstein and Jordan frames
are defined by
\begin{align}
T_{(\tm)\mu\nu} &=
\rho c^2 u_{\mu}u_{\nu}
+p_r k_\mu k_\nu+
p_t\left(g_{\mu\nu}+u_\mu u_\nu-k_\mu k_\nu\right),
\nonumber \\
{\bar T}_{(\tm)\mu \nu} &=
{\bar \rho}c^2 {\bar u}_{\mu}{\bar u}_{\nu}
+{\bar p}_{\bar r}     {\bar k}_{\mu} {\bar k}_{\nu}
+{\bar p}_t
\left({\bar g}_{\mu\nu}
+{\bar u}_{\mu} {\bar u}_{\nu}
-{\bar k}_{\mu} {\bar k}_{\nu}
\right),
\end{align}
where $u^{\mu}$ (${\bar u}^\mu$) and $k^{\mu}$ (${\bar k}^\mu$)
are the four-velocity and unit radial vectors in the Einstein (Jordan)
frame, respectively~\cite{Silva:2014fca}.
Within the first order of Hartle-Thorne's slow-rotation
approximation~\cite{Hartle:1968si}, in the Einstein frame
\begin{align}
\label{def:uk}
u^{\mu}&=
\left(
\frac{1}{\sqrt{-g_{tt}}},0,0,\frac{\Omega}{\sqrt{-g_{tt}}}
\right),
\quad
k^{\mu}&=
\left(
0,\frac{1}{\sqrt{g_{rr}}},0,0
\right), \nonumber \\
\end{align}
and in the Jordan frame
${\bar u}^{\mu}$ and ${\bar k}^{\mu}$
are defined in the same way as Eq.~\eqref{def:uk} with an overbar.
The nonvanishing components of the energy-momentum tensors in both frames
are then given by
\begin{subequations}
\label{components}
\begin{align}
&&
 \label{components_a}
T_{(m)t}{}^t=-\rho c^2,
\quad
T_{(m)r}{}^r=p_r,
\quad
T_{(m)\theta}{}^{\theta}
=
T_{(m)\phi}{}^\phi=p_t,
\\
&&
 \label{components_b}
{\bar T}_{(m)t}{}^t
=-{\bar \rho} c^2,
\quad
{\bar T}_{(m)\bar r}{}^{\bar r}
={\bar p}_{\tilde r},
\quad
{\bar T}_{(m)\theta}{}^{\theta}
={\bar T}_{(m)\phi} {}^\phi={\bar p}_t,
\end{align}
\end{subequations}
and
\begin{subequations}
\label{tphi}
\begin{align}
T_{(m)\phi}{}^t&=
\left(\rho  +\frac{p_t}{c^2}\right) e^{-\nu}\omega r^2\sin^2\theta,
\label{tphi_a}
\\
{\bar T}_{(m)\phi}{}^t
&=
\left(
\bar \rho +\frac{\bar p_t}{c^2}
\right) e^{-\bar \nu}\bar \omega {\bar r}^2\sin^2\theta.
\label{tphi_b}
\end{align}
\end{subequations}
In the Jordan frame, we then make a coordinate transformation
from ${\bar x}^{\mu}=(t,\bar r, \theta,\phi)$ to $x^\mu=(t, r, \theta,\phi)$,
such that
\begin{eqnarray}
{\tilde T}_{(m)\mu}{}^{\nu}
\coloneqq
\frac{\partial {\bar x}^\rho }{\partial x^\mu}
\frac{\partial x^\nu}{\partial {\bar x}^{\sigma}}
 {\bar T}_{(m)\rho}{}^{\sigma}.
\end{eqnarray}
Introducing the components of the energy-momentum tensor
${\tilde T}_{(m)\mu\nu}$ as \eqref{components_a}-\eqref{components_b} with
a tilde, we find
\begin{equation}
\label{bartilde}
{\bar \rho}={\tilde \rho},
\quad
{\bar p}_{\bar r}={\tilde p}_r,
\quad
{\bar p}_{t}={\tilde p}_t,
\end{equation}
and consequently
\begin{equation}
\label{tphi2}
{\bar T}_{(m)\phi}{}^t
=
{\tilde T}_{(m)\phi}{}^t.
\end{equation}
The components of the energy-momentum tensor in the Einstein and Jordan frames are related by~\eqref{relation} and
\begin{eqnarray}
\label{j}
T_{(m)\phi}{}^t&=&A^4 (\vp) \sqrt{\chi} {\tilde T}_{(m)\phi}{}^t.
\end{eqnarray}
Substituting~\eqref{relation}, \eqref{f1}, \eqref{f3}, \eqref{tphi_a}, \eqref{tphi_b},
\eqref{bartilde} and~\eqref{tphi2} into~Eq. \eqref{j} we find
\begin{eqnarray}
\label{omega_inv}
\bar \omega=\omega.
\end{eqnarray}
Thus from~\eqref{f4},
\begin{eqnarray}\
\label{Omega_inv}
\bar \Omega=\Omega.
\end{eqnarray}

The angular momenta in the Einstein and Jordan frames are given by
\begin{eqnarray}
J&=&
\int dr d\theta d\phi \,
r^2 \sin\theta
e^{\frac{\nu+\lambda}{2}} T_{(m)\phi}{}^t,
\\
{\bar J}
&=&\int d\bar{r} d\theta d \phi \,
 \bar{r}^2 \sin\theta
 e^{\frac{\bar \nu+\bar \lambda}{2}} {\bar T}_{(m)\phi}{}^t.
\end{eqnarray}
Using again \eqref{f1}, \eqref{f2}, \eqref{f3}, \eqref{tphi2} and \eqref{j},
we find that the angular momentum is disformally invariant
\begin{eqnarray}
\label{J_inv}
{\bar J}=J.
\end{eqnarray}
From Eqs.~\eqref{Omega_inv} and \eqref{J_inv} we find that the moments of inertia
in the Einstein and Jordan frames, ${I}={J}/{\Omega}$ and
${\bar I}={\bar J}/{\bar \Omega}$, are also disformally invariant
\begin{eqnarray}
{\bar I}=I.
\end{eqnarray}
Thus all quantities associated with rotation are disformally invariant.
Our arguments in this appendix can be applied to a generic class of the Horndeski
theory connected by the disformal transformation~\cite{Bettoni:2013diz}.
%

%

\end{document}